# Evaluation of local stress state due to grain-boundary sliding during creep within a crystal plasticity finite element multi-scale framework

Markian P. Petkov[†], Elsiddig Elmukashfi, Edmund Tarleton, Alan C.F. Cocks

*Department of Engineering Science, University of Oxford, OX1 3PJ, UK*

## ABSTRACT

Previous studies demonstrate that grain-boundary sliding could accelerate creep rate and give rise to large internal stresses that can lead to damage development, e.g. formation of wedge cracks. The present study provides more insight into the effects of grain-boundary sliding (GBS) on the deformation behaviour of realistic polycrystalline aggregates during creep, through the development of a computational framework which combines: i) the use of interface elements for sliding at grain boundaries, and ii) special triple point (in 2D) or triple line (in 3D) elements to prevent artificial dilation at these locations in the microstructure with iii) a physically-based crystal plasticity constitutive model for time-dependent inelastic deformation of the individual grains. Experimental data at various scales is used to calibrate the framework and compare with model predictions. We pay particular consideration to effects of grain boundary sliding during creep of Type 316 stainless steel, which is used extensively in structural components of the UK fleet of Advanced Gas Cooled Nuclear Reactors (AGRs). It is found that the anisotropic deformation of the grains and the mismatch in crystallographic orientation between neighbouring grains play a significant role in determining the extent of sliding on a given boundary. Their effect on the development of stress within the grains, particularly at triple grain junctions, and the increase in axial stress along transverse boundaries are quantified. The article demonstrates that the magnitude of the stress along the facets is highly-dependent on the crystallographic orientations of the neighbouring grains and the relative amount of sliding. Boundaries, transverse to the applied load tend to carry higher normal stresses of the order of 100-180 MPa, in cases where the neighbouring grains consist of plastically-harder crystallographic orientations. An approach to estimate the amount of sliding as a function of GB particle morphology and diffusive accommodation is developed and employed. The possible implications of these findings on the initiation of creep damage are commented on.

## NOMENCLATURE

| | |
|---|---|
| $b$ | Burger's vector |
| $C_{ij}$ | Elastic stiffness matrix of anisotropic crystal |
| $d_p, D, D_b, D_i, D_L$ | Intergranular particle diameter; Grain size; Grain boundary, interface and lattice diffusion |

[†] Corresponding author. M. Petkov; Tel.: +44 (0) 1865 283245; Department of Engineering Science, University of Oxford, OX1 3PJ, UK. Email address: markian.petkov@eng.ox.ac.uk
E. Elmukashfi: elsiddig.elmukashfi@eng.ox.ac.uk
E. Tarleton: edmund.tarleton@eng.ox.ac.uk
A.C.F. Cocks: alan.cocks@eng.ox.ac.uk

| | |
|---|---|
| $E, E_{ij}$ | Young's modulus; Macroscopic strain tensor |
| $\Delta F_0, f_{VA}$ | Activation energy for bypassing; Particle volume per unit grain-boundary area ratio |
| $G$ | Shear modulus |
| $j_s, j$ | Fitting parameters - hardening |
| $k$ | Boltzmann constant |
| $L_d, L_p, L_s$ | Spacing of dislocation junctions, precipitates and solute dispersion atoms |
| $\Delta L_r$ | Annihilated dislocation segment length |
| $N_d, N_0$ | Current and initial number density of dislocation junctions |
| $p, q, r_p^m$ | Flow rule exponents; Intragranular particle radius |
| $t, T, T_m, T_{tk}, T_n$ | Time; Temperature, melting temperature; Tangential, normal and reference traction |
| $v, V$ | Poisson's ratio; Volume |
| $W_c$ | Fitting parameter – static recovery |
| $\alpha_0, \alpha_d, \alpha_p, \alpha_s$ | Obstacle bypassing strength; dislocation junction, precipitate, solute strengths |
| $\alpha^p, \beta^p$ | Parameter groupings |
| $\Gamma^*$ | Fraction of deformation due to grain-boundary sliding |
| $\dot{\varepsilon}_0, \dot{\varepsilon}_e$ | Reference and effective strain rate |
| $\lambda_p, \Lambda$ | Spacing of intergranular particles; Diffusive length |
| $\sigma_{ij}, \Sigma$ | Stress tensor; macroscopic stress tensor |
| $\tau, \tau_{cr}, \bar{\tau}_m^r, \tau^*$ | Resolved and critical resolved shear stress; Type III residual stress; Maximum particle stress |
| $\tau_d, \tau_p, \tau_s$ | Internal resistance - dislocation, precipitates, solutes |

# 1. INTRODUCTION

## *1.1 MOTIVATION*

Grain-boundary sliding is a phenomenon which influences deformation, damage development and failure of polycrystalline materials at high temperatures. The process contributes to inelastic strain accumulation during creep. More importantly, grain-boundary sliding accelerates damage development along grain boundaries, giving rise to intergranular cracks, ultimately resulting in brittle creep fracture [1]. Brittle creep fracture in high-temperature structural components is a major concern in the power generation industry. The constitutive models, employed in industrial structural assessment procedures (e.g. UK's R5 Procedure [2]) or design codes, are often simplistic and they do not take into account the effects of grain-boundary sliding on creep deformation nor any potential increase in stresses along grain boundaries as a result of sliding. However, physically-based multi-scale models have been used in the past to inform the simplistic models employed in industry in capturing important aspects across a range of deformation phenomena. One such multi-scale model, which captures both the dislocation-obstacle interactions at the slip system level and stress redistribution processes in polycrystalline materials, has been developed recently by Hu and Cocks [3–5] and extended by Petkov et al. [6,7]. Model predictions of macro- and mesoscopic creep deformation are in agreement with experiments on austenitic Type 316 stainless steel under a range of loading histories typical of those experienced in high-temperature power plants (e.g. [5,7]). However, this model does not yet account for the contribution to deformation from grain-boundary sliding (GBS). The present article extends the modelling framework to assess how sliding influences the creep response, focusing specifically on how it affects the local distribution of stress along grain facets. Type 316 stainless steel is examined. Before identifying suitable approaches to extend the modelling framework, it is useful to consider how GBS influences the creep response of Type 316 stainless steel.

## *1.2 EXPERIMENTAL OBSERVATIONS OF GRAIN-BOUNDARY SLIDING IN TYPE 316 STAINLESS STEEL*

The present section provides context for the model with respect to the effects of grain-boundary sliding on the global and local deformation of Type 316. Garofalo et al. [8] examined the effects of grain-boundary sliding on the creep deformation of solution-treated (ST) Type 316 stainless steel of grain size 90 μm at 704 and 830°C. The fraction of the total creep strain $E^c$ due to sliding $E^{gbs}$ was quantified in terms of the parameter $\Gamma^* = E^{gbs}/E^c$. The value of $\Gamma^*$ increases with decreasing stress, with values of 6% at 100 MPa and 47% at 41 MPa [8]. This trend does not appear to change with temperature [8,9]. Gates [10,11] and Gates and Horton [12] later examined the effect of stress, creep strain and grain size on $\Gamma^*$ in solution treated (ST) and laboratory-aged (LA) Type 316. This was in order to assess the sensitivity of sliding to the presence of intergranular particles. Larger values of $\Gamma^*$ were recorded for ST (~ 25%) as compared to LA samples (~ 10%) at 40 MPa and 800°C [10]. This indicated that intergranular $M_{23}C_6$ (M is typically Cr) precipitates that form during aging inhibit sliding. Detailed experiments on the formation of wedge cracks in Type 316 by Morris and Harries [9] and Chen and Argon [13] provide insights into the relationship between creep microcracking and grain-boundary sliding. Although the mechanisms proposed for wedge cracking differ in these studies (i.e. athermal decohesion [9] versus accelerated cavitation [13]) each study recognises that grain-boundary sliding enhances stress in the vicinity of a triple point. These studies do not, however, provide details of the local stress state and how this is influenced by the crystallographic orientation of the neighbouring grains and the detailed properties of the grain boundaries. The authors are unaware of experimental or modelling studies in the literature which quantify this stress increase in realistic polycrystalline materials under GBS conditions. Models of the effect of grain-boundary sliding are available in the literature, but these generally employ phenomenological models and simple grain structures.

*1.3 REVIEW OF MODELLING APPROACHES FOR GRAIN-BOUNDARY SLIDING DURING CREEP*
Intrinsic models that describe the micromechanical processes that control the rate of grain-boundary sliding have been developed by Ashby [14]. Ashby demonstrated that in alloys containing populations of intergranular particles, the grain-boundary viscosity is a function of particle morphology, dislocation activity and the dominant diffusion mechanisms in the vicinity of the particles. This study provides limited information, however, about the stress increase along grain boundaries. By contrast, extrinsic models for grain-boundary sliding quantify the contribution from sliding to the overall creep strain in terms of a stress enhancement factor, *f*. Numerical studies using the finite element (FE) method by Crossman and Ashby [15], Hsia et al. [16] and Ghahremani [17] estimate the magnitude of *f* for hexagonal arrays of power-law creeping grains in 2D. Rodin and Dib [18] found higher values of *f* in an idealized 3D array of grains. Onck and van der Giessen [19] investigated the effects of sliding in 2D irregular hexagonal grain arrays on the stress enhancement factor and the stress acting on facets which are approximately normal to the macroscopic maximum principal stress. Higher transverse facet stresses were determined, as well as higher magnitudes of the stress enhancement factor, when microstructural variations are present in the array. Anderson and Rice [20] extended an approach by Needleman and Rice [21] in order to study the effect of grain-boundary sliding on intergranular creep damage in idealised 3D polycrystalline aggregates. The study in [20], however, does not provide insights into the characteristic local stress profile along grain boundaries as a result of sliding.

More elaborate numerical approaches have been developed to model the grain-boundary response in polycrystalline materials, either using specially-developed FE formalisms, e.g. [22], or interface elements, e.g. [23,24]. Adopting such multi-scale modelling approaches has the potential of capturing the global response of the material, while also capturing the local deformation within the microstructure. One set of important features in the microstructure during grain-boundary sliding are the boundary intersections (grain junctions). Pan and Cocks [25] developed FE formulations for superplastic deformation and grain structure evolution by diffusional mass transport along the grain boundaries in the limit that the boundaries slide freely. The study demonstrated that, in 2D quadruple grain junctions are unstable. These redistribute to form triple grain junctions according to out-of-balance grain-boundary tensions and cause the boundaries to have curved profiles [26]. Pan and Cocks [25] impose conservation of mass at each triple by requiring that there is no net flux of material into a triple point. Wilkening et al. [27] extended the approach of [25] by using techniques from semigroup theory. They identified that stress concentrations at triple grain junctions can potentially play a role in void nucleation at these locations. However, neither of these two studies considered the coupling between diffusion-driven grain-boundary sliding and thermally-activated dislocation creep within the grains. A detailed study of superplastic deformation in creeping polycrystals by Bower and Wininger [28] addressed this deficiency. Grain deformation by dislocation-creep, diffusion-driven grain-boundary sliding and grain-boundary diffusion (Coble) creep were simulated within a 2D finite element (FE) model in [28]. The study employed an advancing front algorithm for generation of adaptive meshes which captures the evolution of grain geometries and provides a comprehensive analysis of microstructural evolution in polycrystalline materials during superplastic flow. An important contribution of this study is the introduction of a constraint at the triple junctions. This constraint i) satisfies mass conservation (similar to [25]), while also ii) requiring equilibrium of grain-boundary tensions at these locations, and iii) constraining the migration of boundaries to have a common value at the intersections. It should be noted that the constraint from [28] does not impose any constraint on potential volumetric opening of the triple point due to the relative sliding (tangential) deformation of the boundaries. A study by Wei et al. [29] considered grain-boundary sliding during stress-driven diffusion. The analysis of grain-boundary stresses in [29] during diffusive creep transients for simplified bicrystal configurations revealed a load-transfer process from boundaries of high diffusivities to those of lower diffusivities. Wei et al. [30] extended the framework and found a transition from grain-boundary sliding and diffusion-dominated creep for small grain sizes at low strain-rate, towards dislocation-creep-dominated flow at larger grains sizes and high strain-rates. None of the reviewed studies provides in-depth understanding of the stress, carried by grain boundaries, and its dependence on the crystallographic orientations of the surrounding grains in the polycrystal. Detailed insights into these aspects are required when identifying potential sites for the formation of intergranular cavities and understanding better the underlying physics of these processes.

*1.4 PURPOSE OF THE PRESENT STUDY AND APPROACH*
The main objectives are to quantify and provide an in-depth understanding of:

1) The evolution of local stress fields in the vicinity of grain boundaries and how this is influenced by sliding viscosity and the crystallographic orientation of the neighbouring grains;
2) Physical mechanisms for accommodation of grain-boundary sliding in this alloy.

To achieve this, we develop a polycrystalline, quasi-3D FE framework, which combines physically-based creep models for elastic-plastic-creep deformation of anisotropic grains and grain boundaries. Linear-viscous sliding models are implemented using interface elements within the FE framework in Section 2. When implementing models for sliding, it is important to satisfy the compatibility at grain-boundary intersections such that artificial pores are not created. This is prevented here by imposing constraints at the triple junctions as described in Section 3, which is an additional constraint to those imposed in [25,28]. Here, the grain deformation is modelled using the micromechanical CPFE model described by Petkov et al. [7]. A schematic of the full framework is shown in Fig. 1. We simulate the creep deformation of polycrystalline Type 316 stainless steel, as a common high-temperature structural material, in the presence of grain-boundary sliding and compare the predictions to experimental data at various scales. Note that this physically-based CPFE model considers the combined effects of hardening and thermal recovery of the dislocation structure during creep and allows for accurate representation of the creep response of Type 316 stainless steel. This is in contrast to the empirical constitutive laws employed in [28,30], which simply consider hardening of the dislocation structure during creep. Coble creep and grain growth are not considered here, as there is limited contributions from these processes in Type 316 stainless steel under the temperature and stress range of interest here (500-650°C; 100-300 MPa) [31]. The developed framework is verified for idealised arrays of power-law creeping hexagonal grains by comparing the predictions with results from the literature [17,19] (Section 4). In Section 5, regular hexagonal arrays of grains with different crystallographic orientations which deform by the CPFE model are analysed. The effect of grain neighbours and grain-boundary sliding on the local stress state within the grains and across the boundaries is studied. This provides initial insight into the analysis of more realistic irregular grain structures in the presence of sliding, presented in Section 6, which provides important insights into and quantifies the variation of stress along boundaries and how this is sensitive to the sliding viscosities and relative crystallographic orientations of the neighbouring grains. Experimental data at various scales is used to calibrate the framework and compare with model predictions. The role of intergranular $M_{23}C_6$ precipitates on the sliding response is also briefly examined in Section 7 using a micromechanical model for the sliding velocity. Observations from Section 6 provide important insights into local features of the deformation process, which are relevant to the micro-scale process of cavitation discussed in Section 8.

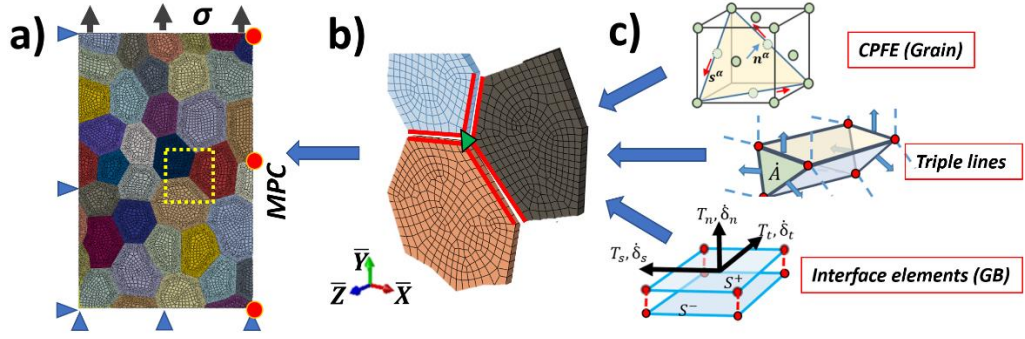

Figure 1. a) The meshed quasi-3D, irregular grain shape aggregate. Regions of the grains, triple grain junctions and boundaries are illustrated in b) with computational models for each structural entity in c).

## 2. CONSTITUTIVE RESPONSE OF GRAINS AND GRAIN BOUNDARIES

In this section, we describe the crystal plasticity model for the grain deformation and the linear viscous model for grain-boundary sliding, and their numerical implementation. Consider a body of volume $V$ and surface $S$, which contains interfaces/grain boundaries of area $\Gamma_c$. The solution of a boundary value problem in which the body is discretized into a number of finite continuum and interface elements with nodal displacements $\mathbf{U}$ is given by the system of equations [32]

$$\mathbf{\Psi} = \int_V \mathbf{B}^T \boldsymbol{\sigma} dV + \int_{\Gamma_c} \mathbf{B}_c^T \mathbf{T}_c d\Gamma - \mathbf{F} = 0 \tag{1}$$

where $\mathbf{B}$ and $\mathbf{B}_c$ are matrices that relate the local strain and separation ($\boldsymbol{\delta}$) across an interface to the nodal displacements, respectively; $\boldsymbol{\sigma}$ is the local stress, $\mathbf{T}_c$ are the interface tractions and $\mathbf{F}$ is a vector of nodal forces. Employing creep models for the bulk response and interface models for sliding in Eq. 1 gives rise to a set of non-linear equations to solve. An iterative procedure for the solution at the end of each time increment via a Newton-Raphson method is used, such that if $\mathbf{U}^i$ is the solution after iteration $i$, a more accurate solution is given as $\mathbf{U}^{i+1} = \mathbf{U}^i - \left[\partial\mathbf{\Psi}^i/\partial\mathbf{U}^i\right]^{-1}\mathbf{\Psi}^i$, where

$$\left(\partial\mathbf{\Psi}/\partial\mathbf{U}\right) = \int_V \mathbf{B}^T \left(\partial\boldsymbol{\sigma}/\partial\boldsymbol{\varepsilon}\right) \mathbf{B} dV + \int_{\Gamma_c} \mathbf{B}_c^T \left(\partial\mathbf{T}_c/\partial\boldsymbol{\delta}\right) \mathbf{B}_c d\Gamma \tag{2}$$

The commercial software ABAQUS is used to obtain solutions to Eq. 1, providing standard shape functions $\mathbf{B}$ for a range of continuum elements. The continuum constitutive response and $\boldsymbol{\sigma}$ and $\partial\boldsymbol{\sigma}/\partial\boldsymbol{\varepsilon}$ are determined by the CPFE model. The framework is not presented for brevity, see [7] for details. Note that for the verification of the modelling framework presented in Section 4, the CPFE model of [7] is replaced by a standard isotropic power-law creep model. In previous studies, model predictions in the absence of grain-boundary sliding using this CPFE model are in agreement with experiments on lattice strains [3,33,34] and macroscopic deformation under short- and long-term loading histories of both ex-service (EX) [5,6] and solution treated (ST) Type 316 [4]. A brief overview of the micromechanical model as applied to Type 316 is presented below. A full description can be found in [3–7,35].

The plastic strain rate in a crystal is determined by summing the slip rates $\dot{\gamma}$, on individual slip systems ($\alpha$) due to thermally-activated dislocation glide at temperature $T$, which are given by [36]

$$\dot{\gamma}^{(\alpha)} = \dot{\gamma}_0 \exp\left(-\left(\Delta F_0/kT\right)\left(1 - \left|\left(\tau^{(\alpha)} + \bar{\tau}_m^r\right)/\tau_{cr}^{(\alpha)}\right|^{3/4}\right)^{4/3}\right) \text{sgn}(\tau^{(\alpha)} + \bar{\tau}_m^r) \tag{3}$$

where $\Delta F_0 = \alpha_0 G_0 b^3$ ($G_0$: shear modulus at 0 K) is the required free activation energy, $k$ is Boltzmann's constant, $\bar{\tau}_m^r$ is the Type III (micro-) internal residual stress [33], $\dot{\gamma}_0$ a reference shear strain rate. The model captures the contributions to the internal resistance $\tau_{cr}$ of the three dominant obstacles to dislocation motion in the alloy - dislocation junctions ($\tau_d$) with spacing $L_{di}$ along the *i*-th slip plane, intragranular precipitates ($\tau_p$) with spacing $L_p$, and discrete solute atoms ($\tau_s$) with spacing $L_s$:

$$\tau_{cr} = \left( \sqrt{\left(\alpha_d G b L_{di}^{-1}\right)^2 + \left(\alpha_p G b L_p^{-1}\right)^2} \right) + \alpha_s G b L_s^{-1} \tag{4}$$

where $G$ is the shear modulus, $b$ is the Burger's vector, and $\alpha_p$, $\alpha_d$ and $\alpha_s$ are dimensionless obstacle strength parameters [37]. Grain-size effects are not considered in the model. $L_{di}$ is a function of the dislocation-junction obstacle density, $N_{di}$ ($\dot{L}_{di} = -N_{di}^{3/2} \dot{N}_{di} / 2$) due to self- and latent-hardening as a result of dislocation multiplication related to the increment in the resolved shear strain $\Delta\gamma$ on a slip plane:

$$\Delta N_{di}^{self} = j_s \Delta\gamma_i, \quad \Delta N_{di}^{latent} = j \sum_{k \neq i} \Delta\gamma_k \tag{5,6}$$

where $j_s$ and $j$ are hardening parameters, $i = 1$~$4$ and $\Delta\gamma_i$ is the resolved shear strain increment from all three slip systems for the *i*-th slip plane [3]. Dynamic recovery is captured through a model introduced by Kocks [38], which is applied at the slip system level as described in [6]. Dynamic recovery is a function of the probability that a dislocation segment length ($\Delta L_r$) is annihilated for a given increment of the resolved shear strain $\Delta\gamma$ according to

$$\Delta N_{di} = -\Delta L_r N_{di} \Delta\gamma_i / b \tag{7}$$

Thermal recovery of the dislocation network is modelled by assuming that at elevated temperature long dislocation links grow at the expense of shorter ones, as described in [5]. As a result, the decrease in dislocation junction number density due to thermal recovery can be expressed as

$$\Delta N_{di} = -\left(2 W_c D_c G b^5 / kT\right) N_{di}^3 \Delta t \tag{8}$$

where $D_c$ is the core diffusivity and $W_c$ is a dimensionless parameter controlling thermal recovery [35].

The interface element is implemented as a User Element (UEL) in ABAQUS, following the approach, described by Elmukashfi and Cocks [32], which adopts a surface-type interface formulation from [39]. Simpler and more versatile 8-noded linear interface elements in 3D are used in this study, compared to the 9-noded quadratic 2D interface elements adopted in [28]. A full description of the interface element approach employed in the present article can be found in [32]. The formulation for 3D problems is given in Appendix A. For the constitutive grain-boundary response, a traction-separation rate law is assumed which decomposes into normal ($\dot{\delta}_n$) and tangential ($\dot{\delta}_{tk}$) separation rates:

$$\dot{\delta}_n = \dot{\delta}_n^e = \dot{T}_n / K_n \tag{9}$$

$$\dot{\delta}_{tk} = \dot{\delta}_{tk}^e + \dot{\delta}_{tk}^{cr} = \left(\dot{T}_{tk} / K_t\right) + \dot{\delta}_{0t} \left(T_{tk} / \sigma_0\right) \tag{10}$$

where $\dot{\delta}_i^e$ and $\dot{\delta}_i^{cr}$ are the elastic and viscous components, respectively, and $k = 1,2$ are the two tangential directions. $K_n$ and $K_t$ are the normal and tangential stiffnesses and $\dot{\delta}_{0t}$ is the rate of sliding at a tangential traction $T_{tk} = \sigma_0$. The elastic response is chosen to be sufficiently stiff so that it has a negligible effect on the material response. Note that a computational scheme for updating the interface element configuration within a large deformation analysis is employed throughout this study, which uses shape functions for the interface element in the deformed configuration. This updating framework is briefly presented in Appendix B. Implementation of the framework described in Appendix A does not automatically satisfy compatibility where a number of interface elements meet, i.e. triple points in 2D or triple lines in 3D. In the following section, a constraint at the triple junctions is introduced to ensure that the compatibility condition is met. This is implemented through a penalty function. The employed triple junction constraint is an additional constraint to those in [28]. This is described in detail in the following section.

## 3. MODELLING OF TRIPLE GRAIN JUNCTIONS DURING SLIDING

In this study, thin-sliced aggregates, one element thick, with irregular grain shapes under quasi-3D plane strain boundary conditions are considered. This is in an attempt to reduce simulation times, while providing sufficient information on stress and strain fields and how these depend on crystallographic orientation. Consider the interface element of thickness $h$ in Fig. 2.a). Where interface elements meet at a triple point, a triangular region $N$ of sides $h_{12}, h_{13}, h_{23}$, shown in Fig. 2.b), is created. The rate of change of area of any triangular region $N$ can be determined by integrating the normal component of velocity over its perimeter:

$$\dot{A}_\Delta^s = \oint \dot{u}_i^s n_i ds \tag{11}$$

where $n_i$ is the outward unit vector to a side of the element and $\dot{u}_i^s$ is the respective displacement rate in the sliding direction of the adjacent interface element. A linear variation of $\dot{u}_i^s$ is assumed between the nodal values. Applying Eq. 11 to a general 2D triple point element of arbitrary orientation and dividing through by $h_{13}$ gives a rate quantity, independent of the boundary thickness:

$$\dot{L}_\Delta^s = \frac{\dot{A}_\Delta^s}{h_{13}} = \frac{1}{2}\left[\left(\dot{u}_i^{s1} + \dot{u}_i^{s3}\right)n_i^{13} + \left(\dot{u}_i^{s1} + \dot{u}_i^{s2}\right)n_i^{12}\frac{\sin\alpha_\Delta}{\sin\gamma_\Delta} + \left(\dot{u}_i^{s2} + \dot{u}_i^{s3}\right)n_i^{23}\frac{\sin\beta_\Delta}{\sin\gamma_\Delta}\right] \tag{12}$$

where angles are defined in Fig. 2. When the triple point is in equilibrium, such as in a hexagonal array of grains, region $N$ is an equilateral triangle. Eq. 12 then becomes

$$\dot{L}_\Delta^s = \frac{1}{2}\left[\left(\dot{u}_i^{s1} + \dot{u}_i^{s3}\right)n_i^{13} + \left(\dot{u}_i^{s1} + \dot{u}_i^{s2}\right)n_i^{12} + \left(\dot{u}_i^{s2} + \dot{u}_i^{s3}\right)n_i^{23}\right] \tag{13}$$

The constraint requires $\dot{L}_\Delta^s = 0$ (i.e. $\dot{A}_\Delta^s = 0$) which is implemented through the use of penalty functions. Eq. 12 can be re-written as $\dot{L}_\Delta^s = \mathbf{A}\dot{\mathbf{U}} = 0$, where the matrix $\mathbf{A}$ for a general triple point element is given in Appendix C. A prescribed penalty function, $P$, is imposed on $\dot{L}_\Delta^s$ according to

$$\sum_\Delta \delta\dot{L}_\Delta^s P\dot{L}_\Delta^s = \delta\mathbf{U}^T\sum_\Delta \mathbf{A}^T P\dot{L}_\Delta^s = \delta\mathbf{U}^T\sum_\Delta \mathbf{A}^T S_\Delta^s = 0 \tag{14}$$

The quantity $S_\Delta^s = P\dot{L}_\Delta^s$ has the dimensions of force and summation is over all triple points within the body. In a quasi-3D, thin-slice the triple points become triple lines and the triple point element becomes a prism (Fig. 2.c). The prism can change in length, but the constraint requires its cross-sectional area to remain constant along its length. This is achieved by imposing the constraint from Eq. 14 at the triangles formed by nodes (1-3) and (4-6) in Fig. 2.c). Linear variation of displacement rates is assumed along the axis of the prismatic element. As a result the constraint is satisfied at all points along the element. For this element, the rate quantity $\dot{L}_\Delta^s$ from Eq. 12 becomes

$$\dot{L}_{Line}^s = \frac{1}{2}\frac{\dot{A}_\Delta^s}{h_{13}} + \frac{1}{2}\frac{\dot{A}_\Delta^{s\prime}}{h_{13}'} = \frac{1}{4}\begin{bmatrix}\left(\dot{u}_i^{s1} + \dot{u}_i^{s3}\right)n_i^{13} + \left(\dot{u}_i^{s4} + \dot{u}_i^{s6}\right)n_i^{13} + \\ +\left(\dot{u}_i^{s1} + \dot{u}_i^{s2}\right)n_i^{12}\frac{\sin\alpha_\Delta}{\sin\gamma_\Delta} + \left(\dot{u}_i^{s4} + \dot{u}_i^{s5}\right)n_i^{12}\frac{\sin\alpha_\Delta'}{\sin\gamma_\Delta'} + \\ +\left(\dot{u}_i^{s2} + \dot{u}_i^{s3}\right)n_i^{23}\frac{\sin\beta_\Delta}{\sin\gamma_\Delta} + \left(\dot{u}_i^{s5} + \dot{u}_i^{s6}\right)n_i^{23}\frac{\sin\beta_\Delta'}{\sin\gamma_\Delta'}\end{bmatrix} \tag{15}$$

where the primed quantities are associated with the rear face of the prismatic triple line element. For this element, Eq. 14 takes the same form and the constraint requires

$$\sum_\Delta \mathbf{A}^T S^s_{Line} = 0 \tag{16}$$

where $S^s_{Line} = P\dot{L}^s_{Line}$ with matrix **A** for the triple line element given in Appendix C. For the triple line, the quantity from Eq. 16 is then added to Eq. 1 and a solution to the set of equations

$$\bar{\Psi} = \Psi + \sum_\Delta \mathbf{A}^T S^s_{Line} = \int_V \mathbf{B}^T \boldsymbol{\sigma} dV + \int_{\Gamma_c} \mathbf{B}_c^T \mathbf{T}_c d\Gamma + \sum_\Delta \mathbf{A}^T S^s_{Line} - \mathbf{F} = 0 \tag{17}$$

needs to be determined. The Newton-Raphson scheme for the iterative solution of Eq. 17 is

$$\frac{\partial \bar{\Psi}}{\partial \mathbf{U}} = \frac{\partial \Psi}{\partial \mathbf{U}} + \sum_\Delta \frac{\partial \mathbf{A}^T S^s_{Line}}{\partial \mathbf{U}} = \int_V \mathbf{B}^T \frac{\partial \boldsymbol{\sigma}}{\partial \boldsymbol{\varepsilon}} \mathbf{B} dV + \int_{\Gamma_c} \mathbf{B}_c^T \frac{\partial \mathbf{T}_c}{\partial \boldsymbol{\delta}} \mathbf{B}_c d\Gamma + \sum_\Delta \mathbf{A}^T \frac{\partial S^s_{Line}}{\partial L^s_{Line}} \mathbf{A} \tag{18}$$

Parametric studies in which the penalty function is systematically increased have been carried out to ensure that the linear growth rates ($\dot{L}^s_\Delta$, $\dot{L}^s_{Line}$) are constrained. In Section 5.3, it will be demonstrated that artificial voids can form at triple grain junctions in the absence of the constraint and influence the local stress state. Note that in the calculations presented here interface elements of zero thickness are employed.

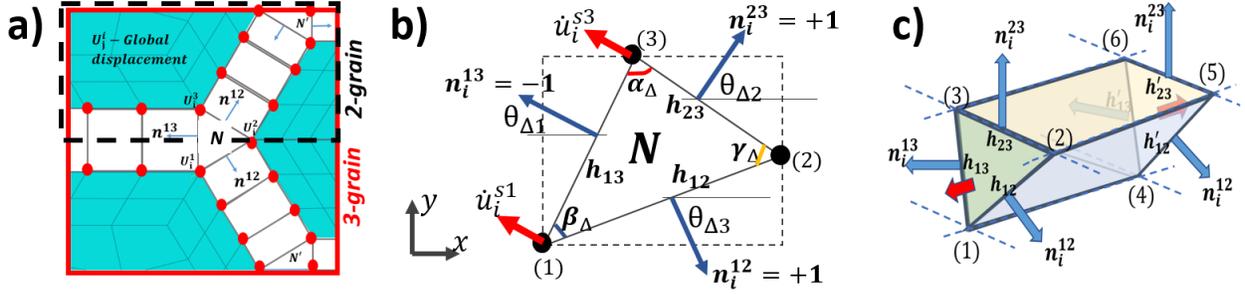

Figure 2. a) Illustration of 2- and 3-grain hexagonal arrays, and formation of triple point at interface element intersections. Schematic of an arbitrary b) triple point (2D) and c) triple line (3D) element.

## 4. VERIFICATION OF FRAMEWORK FOR A POWER-LAW CREEPING MATERIAL

Before employing the framework to study the creep response using the anisotropic model of Section 2, we verify the approach using a simple isotropic constitutive model, for which results are available in the literature [17,19]. The geometry of three hexagonal grains, shown in Fig. 2.a), is modelled in 2D plane strain. The mesh consists of standard plane strain linear 3-noded (designated CPE3 by ABAQUS) and 4-noded bilinear (CPE4) elements for the grains, and 4-noded linear interface elements for the boundaries. The deformation of the grains is described by an isotropic power-law creep model, given by $\dot{\varepsilon}_e = \dot{\varepsilon}_0 (\sigma_e/\sigma_0)^n$. Macroscopic stresses of 144.25 MPa were imposed on the top and right-hand faces of the model resulting in an applied macroscopic effective stress $\Sigma_e$ of 250 MPa. By progressively increasing the relative sliding rate in Eq. 10, i.e. $\dot{\delta}_{0t}$, differences in macroscopic creep rates and steady-state stress fields were recorded for creep exponents $n = 1 - 9$. The parameter $\phi_0 = \dot{\varepsilon}_0 D/\dot{\delta}_{0t}$, describing the ease with which the interface deforms as compared to the creep rate of the grains of size $D$, is introduced. This allows for comparison across configurations and different reference creep rates, $\dot{\varepsilon}_0$. The increase in strain rate with decreasing values of $\phi_0$ can be quantified by the stress enhancement factor $f = (\dot{E}_e(\Phi_0)/\dot{E}_e(\infty))^{1/n}$, where $\dot{E}_e(\Phi_0)$ is the macroscopic effective strain rate for a given value of $\phi_0$ and $\dot{E}_e(\infty)$ is that in the absence of sliding [17]. The variation of $f$ with $n$ and $\phi_0$ is shown in Figs. 3.a) and b). The predictions agree with results for the same configuration from [17,19]. The limit of free sliding is approached for $\phi_0 \ll 1$. The same results were also obtained for this geometry under quasi-3D plane strain, where the mesh consisted of linear 6-noded (wedge) and 8-noded (brick) elements for the grains, and 8-noded linear interface elements for the boundaries (Figs. 3.a-b). It is important to note that values of *f* from the present study are higher due to the use of a finer mesh, compared to the mesh referenced as 'Fine' employed by Ghahremani [17]. Examples of steady-state stress fields in the y-direction, i.e. $\sigma_{yy}$, are illustrated in Figs. 3.c) and d).

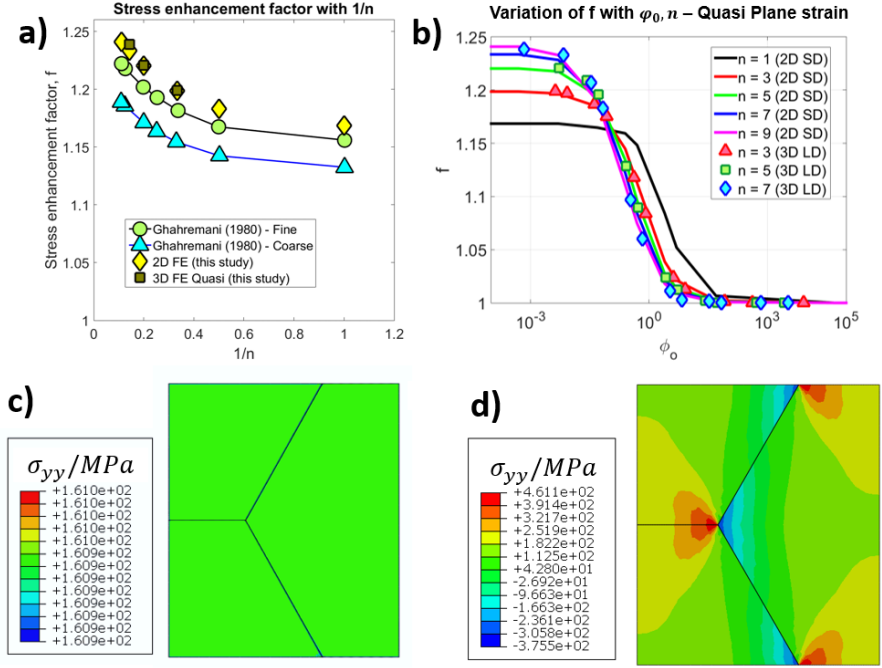

Figure 3. Stress enhancement factor with a) $n$ and b) $\phi_0$ – 2-grain hexagonal grain array, power-law creep. Contour plots of steady-state $\sigma_{yy}$ in 3-grain array, $n = 7$ for a) no sliding and b) free sliding, $\phi_0 = 10^{-3}$.

A heterogeneous stress field is predicted in the presence of sliding, with elevation of tensile or compressive stress on either side of a triple point, while the stress field is uniform in the absence of sliding. The average stress on the transverse boundary was found to be 4 times greater than the average stress on the bulk grain edge in the limit of free sliding, which agrees with [16]. These results provide confidence in the developed methodology to capture the response of creeping polycrystals in the presence of sliding. In the following sections, we analyse the effects of grain-boundary sliding when grains deform by the CPFE model described in Section 2 [7].

## 5. MODELLING REGULAR HEXAGONAL ARRAYS OF GRAINS USING CPFE MODEL

In this section, we analyse the effects of grain-boundary sliding on the behaviour of regular, hexagonal arrays of grains with different crystallographic orientations, as shown in Figs. 4.a-c). The simulations provide a preliminary assessment of the effect of grain-boundary sliding and an indication of the sensitivity of stress state to the imposed boundary conditions. This serves as a precursor to analysing the response of a larger aggregate in Section 6.

### 5.1. CREEP RESPONSE OF ANISOTROPIC ARRAYS OF GRAINS IN THE PRESENCE OF SLIDING

The interface sliding and CPFE models, introduced in Section 2, are implemented as UEL and UMAT routines within ABAQUS. Linear 6- (C3D6) and 8-noded (C3D8) continuum elements are used for the grains; the 8-noded linear interface element described in Appendix A are used to model sliding. A sensitivity study of the interface stiffness ($K_n$, $K_t$) revealed that a value >$10^6$ MPa/mm restricts the elastic opening and tangential separations to magnitudes lower than their creep counterparts. The value of the penalty function, *P*, for enforcing the constraint on the triple line was selected to be 8 x $10^{10}$ N/mm. This constrains the opening of the triple line (Section 5.3) without introducing numerical difficulties. The arrays of Figs. 4.a)-c) are analysed with the different crystallographic orientations shown in Fig. 4.d), where S and S2 are plastically softer orientations than H with respect to the loading direction. The CPFE model is calibrated to experimental short-term plasticity and creep data of Type 316 stainless steel at 625°C (See Table 1).

### 5.1.1. Variation of strain-rate in the presence of sliding

Consider the 3-grain aggregate from Fig. 4.a) loaded in uniaxial tension under quasi-3D plane strain. The reference sliding rate ($\dot{\delta}_{0t}$) is progressively increased until there is limited change of the predicted creep response, which is taken as the free-sliding limit. The characteristic S-shaped curve from Fig. 3.b) for the minimum creep rate $\dot{E}_e$ with decreasing $\phi_0 = \dot{\varepsilon}_0 D/\dot{\delta}_{0t}$ (reference value $\dot{\varepsilon}_0 = \dot{E}_{e,locked}$ in the absence of sliding) is shown in Fig. 5. Values of $\dot{E}_e/\dot{\varepsilon}_0$ are equal to 1.7, 2.85 and 3.6 for $\phi_0$ = 0.4, 0.1 and 0.04, respectively. A mesh convergence study of the variation of macroscopic minimum creep rate with $\phi_0$ is shown in Fig. 5.b). The meshes examined consist of 1896 bulk and 112 interface elements (div4), 2971 bulk and 336 interface elements (div3), and 8097 bulk and 672 interface elements (div2). No significant change in $\dot{E}_e/\dot{\varepsilon}_0$ with mesh refinement is observed and in an attempt to increase computational efficiency, mesh reference "div4" was selected for further simulations. Note that examples of predicted macroscopic creep curves are shown in Fig. 5.a) for the 3-grain SSH configuration, where the transition from "locked" to freely-sliding grain boundaries is observed for $\phi_0 \ll 1$. The minimum creep rate is arbitrarily defined here as the average macroscopic creep rate over the final 10% of the simulation time.

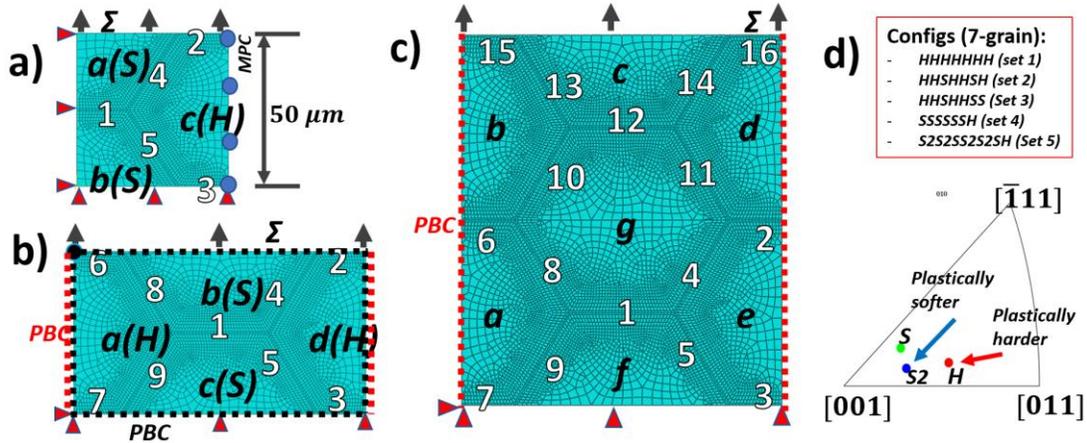

Figure 4. a) 3-grain hexagonal array (ref div4). b) The 4-grain array with periodic boundary conditions imposed on vertical faces. c) The 7-grain array with a "floating" grain in its centre and periodic boundary condition (PBC) imposed on vertical faces. d) Crystallographic orientations used in the analysis of the 7-grain aggregate for grains 'a'-'g'- S and S2 are plastically softer than H.

The 3-grain aggregate provides an initial indication of the effect of sliding on creep rate, but the symmetry boundary conditions are not appropriate for the CP models considered [40]. For the 4- and 7-grain geometries of Figs. 4.b)-c) periodic boundary conditions (PBCs) are employed on the vertical cell faces with the nodal displacements along the right face linked to those along the left face. This is possible when grains at opposite faces have the same crystallographic orientation (e.g. Set 5 in Fig. 4). The PBCs take the form $u_i = \bar{\varepsilon}_{ik} x_k + u_i^*$, where $\bar{\varepsilon}_{ik}$ are the average strains and $u_i^*$ is the periodic part of the nodal displacements, as described by Arora et al. [40]. The response under a uniaxial load of 220 MPa at 625°C for the 3-, 4- and 7-grain aggregates with crystallographic orientations SSH, HSSH and HHSHHSS are compared in Fig. 5.a) for values of $\phi_0$ = +∞, 0.2 and 0.04. The extent of primary creep is greater when the boundaries slide freely. For the assumed model here, the primary creep response is controlled by the evolution of microstructural state (i.e. dislocation structure) and the redistribution of stress from an initial state to a steady state distribution, as demonstrated in previous studies [41]. The results here demonstrate that when boundaries slide freely, the stress redistribution process is more pronounced as the shear stresses supported by the grain-boundaries initially relax, giving rise to a more extensive primary creep response in this limit. When sliding is restricted, the creep responses of the different arrays of Fig. 4 are identical, as shown in Fig. 5.a). However, in the presence of sliding a weaker increase in creep strain rate with decreasing value of $\phi_0$ is observed for the 7-grain aggregate compared to the 3-grain cell, with the 4-grain cell lying in-between the two.

The constraint on the deformation of the internal grains decreases with increasing size of the simulation cell and this observation suggests that the weaker the constraint the slower the increase in creep rate in the presence of sliding. This difference is illustrated in Fig. 5.b) by the variation of normalised strain rate $\dot{E}_e/\dot{\varepsilon}_0$ with $\phi_0$ for the three arrays. The following sections will demonstrate that this effect is driven by the variation of local stress state arising from the different internal constraints.

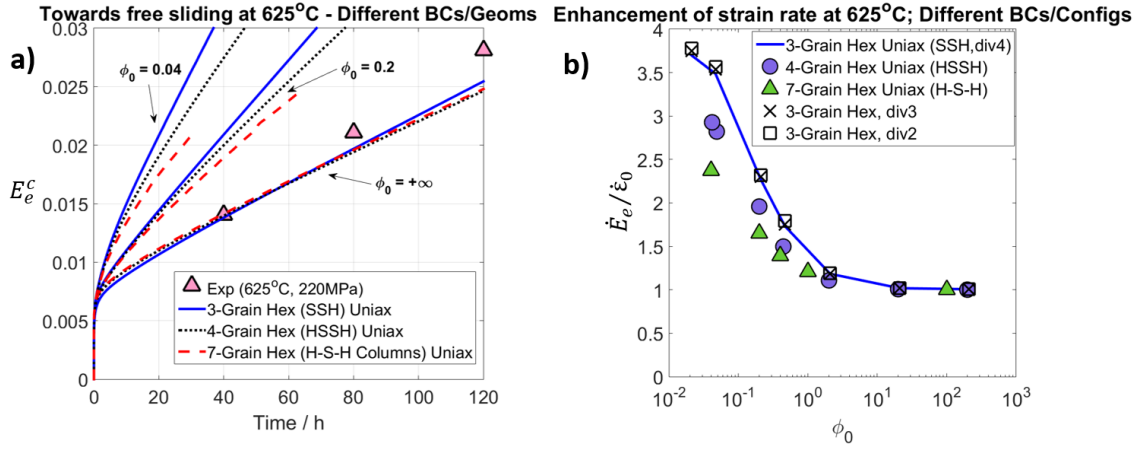

Figure 5. a) Macroscopic creep strain for different values of $\phi_0$ for 3-grain SSH, 4-grain HSSH and 7-grain H-S-H (Set 3) configurations at 625°C under uniaxial loading. b) Variation of $\dot{E}_e/\dot{\varepsilon}_0$ with $\phi_0$ for the three hexagonal geometries from a). For the 3-grain array, div2 is the finest mesh studied with 192 interface elements along an inclined boundary and div4 is the coarsest with 32 interface elements.

### 5.1.2. Contribution of grain-boundary sliding to overall deformation rate

As discussed in Section 1, the enhancement of macroscopic strain rate when grain boundaries slide is due to: i) the resulting non-uniform intragranular stress field; and ii) the direct contribution to the global strain from sliding. A procedure for estimating contribution (ii) is employed in this study using a similar approach to [30]. As will be shown later, the macroscopic components of strain rate and the contributions from sliding can be measured experimentally (see [9]), allowing direct calibration and comparison with the FE analyses. Consider a polycrystal of volume $V$ that contains grain boundaries of total area $\Gamma_c$ that experiences a macroscopic strain rate $\dot{E}_{ij}$ as a result of straining of the grains ($\dot{\varepsilon}_{ij}$) and grain boundary sliding ($v_s$). The Principal of Virtual Power is written as

$$\dot{E}_{ij}\Sigma^*_{ij} = \frac{1}{V}\left[\int_V \dot{\varepsilon}_{ij}\sigma^*_{ij}dV + \int_{\Gamma_c} v_s \sigma^*_{ij} n_i s_j d\Gamma\right] \tag{19}$$

where $\Sigma^*_{ij}$ is an arbitrary applied stress and $\sigma^*_{ij}$ is an arbitrary stress field in equilibrium with $\Sigma^*_{ij}$; $n_i$ is the normal to the sliding boundary and $s_i$ is a unit vector in the direction of sliding. Selecting a uniform stress $\sigma^*_{ij}$, such that $\sigma^*_{ij} = \Sigma^*_{ij}$, allows the components of strain rate arising from deformation of the grains and boundaries to be determined as

$$\dot{E}_{ij} = \frac{1}{V}\left[\int_V \dot{\varepsilon}_{ij}dV + \int_{\Gamma_c} v_s n_i s_j d\Gamma\right] \tag{20}$$

Consider a body subjected to a uniaxial stress $\Sigma_{yy}$ which results in a strain rate $\dot{E}_{yy}$. The fraction of the total macroscopic strain rate in the y-direction due to sliding, $\Gamma^*_{yy}$, is

$$\Gamma^*_{yy} = \dot{E}^{gbs}_{yy}/\dot{E}_{yy} = (1/V\dot{E}_{yy})\int_{\Gamma_c} v_s n_2 s_2 dS \tag{21}$$

Eq. 21 is now used to obtain the values of $\Gamma^*_{yy}$ for the 3-, 4- and 7-grain aggregates. In order to apply Eq. 21 to these geometries, the predicted sliding velocity profiles $v_s(x)$ along individual grain boundaries were numerically integrated. As an example, the steady-state sliding velocity profiles along grain boundaries of the 4-grain aggregate at 625°C are given in Fig. 6.a). These predictions are presented here for illustrative purposes in demonstrating the variation of sliding rate along different boundaries. In Section 6, the predicted sliding displacements and rates along grain boundaries of the more realistic polycrystalline aggregates are compared to experimental data. To compute $\Gamma^*_{yy}$, the creep rates from Fig. 5.a) and the sliding velocity profiles along each boundary (e.g. Fig. 6.a) are used. The computed values of $\Gamma^*_{yy}$ for the 3-grain SSH, 4-grain HSSH and 7-grain H-S-H (Set 3) configurations under uniaxial loading at 625°C and $\phi_0 = 0.04$ are 20.1%, 20.9% and 21.1%, respectively. At this particular value of $\phi_0$, the contributions from grain-boundary sliding to the total deformation rate are virtually the same for the three idealisations. By computing the values of $\Gamma^*_{yy}$ for each of the five 7-grain configurations examined, larger contributions to the overall deformation due to sliding are found in the "softer" aggregates at a given stress level (Fig. 6.b). While indicative, at present the authors were unable to find experimental data in the literature confirming this trend. Section 6 demonstrates how experimental data for $\Gamma^*_{yy}$ can inform the calibration of the interface element, allowing for direct comparison between model predictions and experimental data at macro- and microscopic scales.

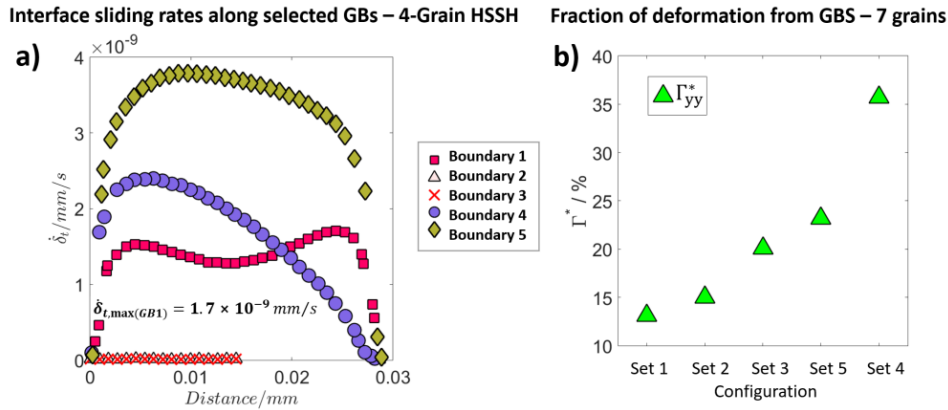

Figure 6. a) Example, steady-state sliding rate along selected boundaries (4-grain HSSH) at $\phi_0 = 4 \times 10^{-2}$. In b), the fractions of deformation due to GBS in y-direction ($\Gamma^*_{yy}$) are shown for the different crystallographic sets.

Another important conclusion from the predictions is that boundaries normal to one of the principal loading axes can also slide (see Boundary 1 in Fig. 6.a). However, the steady-state sliding rates along these boundaries are approximately 1-3 orders of magnitude slower than those along the inclined boundaries. This phenomenon arises from the elastic-plastic anisotropy mismatch between the grains either side of the boundary. Because of this mismatch, the structural problem is not symmetric about the interface. Thus, contrary to the commonly held view, sliding can occur on boundaries normal to the loading axis. This is discussed in more detail in Section 8.

### 5.2. *VARIATION OF STRESS STATE DUE TO SLIDING AND IMPOSED BOUNDARY CONDITIONS*
The major contribution to the increased deformation rate due to sliding arises from the non-uniform intragranular stress field as a result of the relaxation of shear stresses along sliding interfaces [16]. Here, we assess the resulting stress distribution in the absence and presence of sliding in the small grain aggregates. The predicted heterogeneous $\sigma_{yy}$ stress fields within the 4-grain HSSH and 7-grain H-S-H configurations under a uniaxial load of 220 MPa at 625°C are compared for the cases of no sliding and $\phi_0 = 4.0 \times 10^{-2}$ in Fig. 7. The distinct pattern in the plots for small values of $\phi_0$ is similar to that for the power-law creeping material in Section 4. The stress profile shows an elevation of stress in tension or compression across the transverse interfaces with a concentration of stress near the triple junctions. It is important to note that the additional heterogeneity due to anisotropy of each individual grain in the CPFE model enhances this effect. Also, the detailed form of the stress field depends on the crystallographic orientation of the grains either side of the boundary. This is illustrated in the line plots of $\sigma_{yy}$ shown in Fig 8.b) along transverse boundaries 1, 2 and 6 for the 7 grain H-S-H array, where the stresses are much higher between the hard orientations than the softer orientations, agreeing with observation from [42–44]. The use of periodic boundary conditions for the vertical boundaries is appropriate for the H-S-H arrangement, but the use of symmetry conditions at the top and bottom faces is not generally appropriate (apart for some very specific crystal orientations). This can result in spurious stress distributions at these boundaries as illustrated in Fig 8.a), which shows the stress distribution along the equivalent boundaries to Fig 8.b) for the 4-grain array. The stresses along grain boundaries 2 and 6 (which lie along the boundary of the analysed cell) are now different to each other and contain opposing singularities where the boundary meets the bottom corners of the cell. Within half a distance of half a grain diameter from the boundary, however, the two simulation provide equivalent stress distributions (as illustrated by the stress along boundary 1 for the two simulations in Fig 8). We use this observation in section 6 when modelling the behaviour of an irregular array of grains.

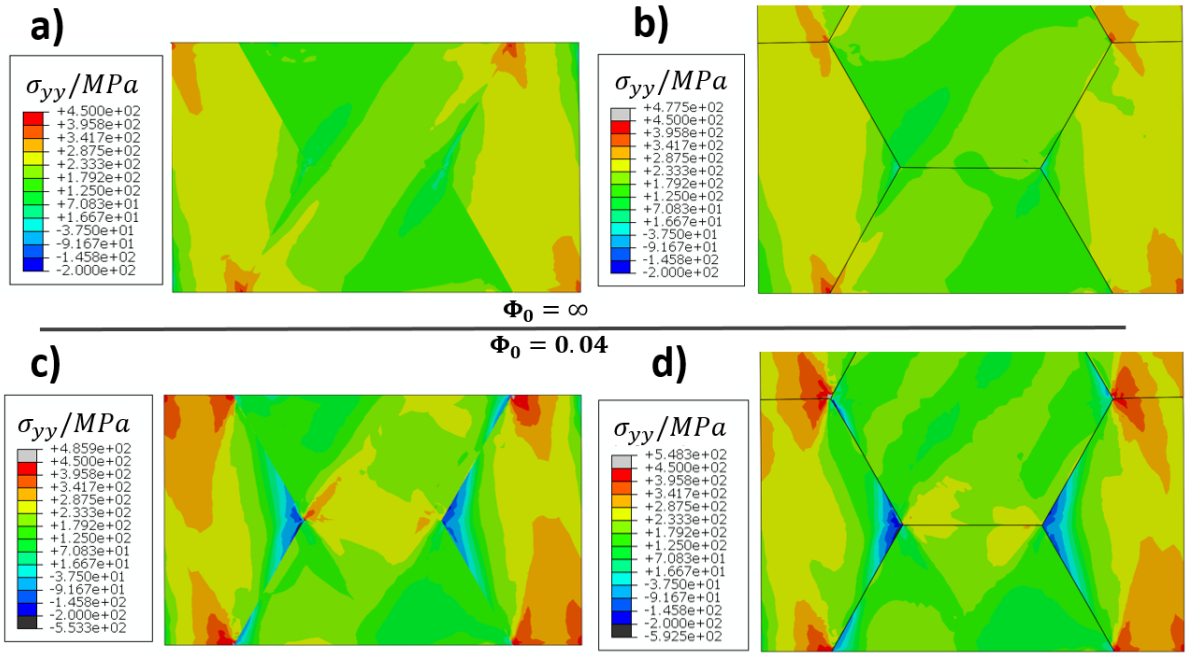

Figure 7. Contour plots of $\sigma_{yy}$ after 50 hours at 625°C in the absence of sliding in a) the 4-grain and b) 7-grain aggregates; and at $\phi_0 = 4 \times 10^{-2}$ in c) and d), respectively. Only the lower portion of the 7-grain cell is shown.

For a power-law material the stress is uniform within an array of grains in the absence of sliding. For an array of anisotropic grains it is not and mild singularities in stress are generated at the triple points, due to discontinuities in structure and properties at these points. These singularities are more pronounced in the presence of sliding. Sliding also enhances the average stress across boundaries normal to the applied stress [65]. For example, for the 7-grain cell, the average value of $\sigma_{yy}$ increases from 178 MPa to 238 MPa across the S-S boundary 1 as $\phi_0$ is reduced from ∞ to $4 \times 10^{-2}$, while it increases from 245 MPa to 315 MPa across the H-H boundaries 2 and 6.

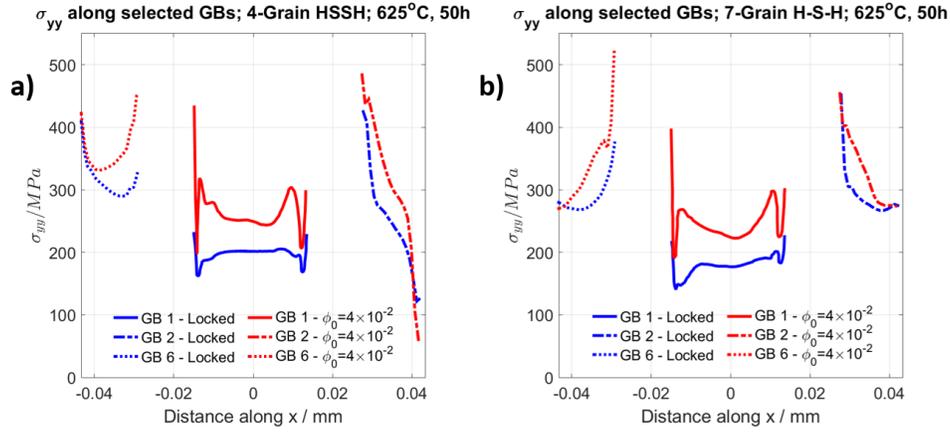

Figure 8. Line plots of $\sigma_{yy}$ under uniaxial loading after 50 hours with distance along x-coordinate for grain boundaries (GBs) 1,2 and 6 within the a) 4-grain HSSH aggregate and b) 7-grain H-S-H columnar configuration (Set 3) at 625°C. Scenarios in the absence of sliding and $\phi_0 = 4 \times 10^{-2}$ are illustrated.

### 5.3. IMPORTANCE OF THE TRIPLE LINE CONSTRAINT

The simulations presented in Section 4 for a power-law creeping material have been conducted with and without implementation of the triple point constraint. Equivalent solutions were obtained. This is due to the strong symmetry of the configuration and the isotropic material model adopted. However, the triple junction constraint plays an important role for the crystal plasticity model. Contour plots of $\sigma_{yy}$ stress fields are shown in Fig. 9 for the CPFE model for a) no-sliding b) freely-sliding boundaries with the constraint and c) freely-sliding boundaries without the constraint. When sliding is absent, the constraint is redundant (Fig. 9.a). In the presence of sliding, the triple junction in Fig. 9.b) rotates due to the anisotropic deformation of each grain and the imposed constraint at the triple point. This is not the case in Fig. 9.c) where a "void" forms in the absence of the constraint, violating the compatibility requirement. Similar observations are made for other combinations of grains and crystallographic orientations considered here. More importantly, Figs. 9.b-c) indicate that when the constraint is absent the local stress concentration at the triple point is modified significantly when sliding occurs. It can be seen that a difference of 50-100 MPa in local axial stress compared to an applied stress of 200 MPa can arise within 0.5-1.0 μm of the triple junction in the absence of the constraint.

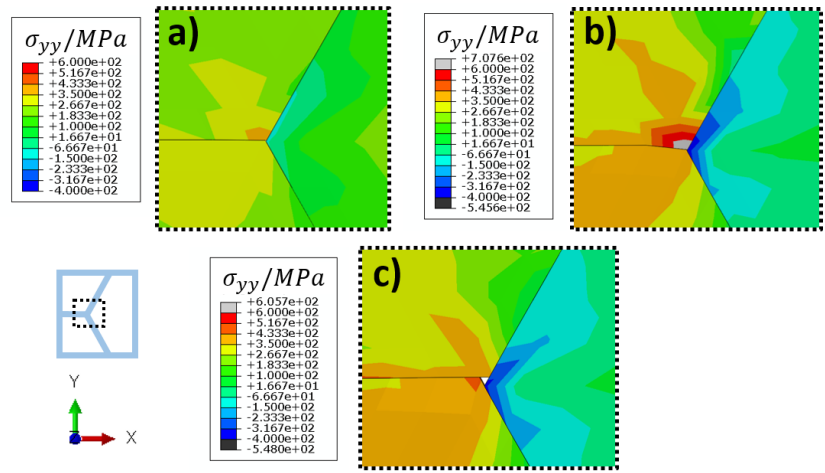

Figure 9. Plots of $\sigma_{yy}$ in a 3-grain cell after creep under uniaxial loading – a) no sliding, b) freely-sliding boundaries with the constraint, and c) freely-sliding boundaries without the constraint (close-up of the triple junctions).

This is supported by [28,45] on the potential inaccuracies in local stress at such locations. This is an important observation and illustrates the importance of impose the correct constraint at the triple point/line. In addition, inconsistencies introduced by omitting the constraint could have a substantial impact on predicting intergranular damage in the presence of grain-boundary sliding. It is worth pointing out that extension of this constraint formulation to full-3D scenarios, where the triple lines from Fig. 2 intersect to form apices, is possible but is outside the scope of this article.

## 6. Modelling of large polycrystalline aggregates

In this section, the CPFE-interface element framework is employed to study the variation of local stress state due to grain-boundary sliding in larger polycrystalline aggregates. The particular case of Type 316 stainless steel at 625°C under an applied uniaxial stress of 220 MPa is simulated. Morris and Harries [9] measured experimentally the fraction of deformation due to sliding as $\Gamma_{yy}^* \approx 6\%$ under these conditions. The CPFE model and interface elements are calibrated to simultaneously capture the macroscopic creep deformation of the material and also give $\Gamma_{yy}^* \approx 6\%$ (see later). This allows calibration with experimental data extracted from both macroscopic (e.g. creep curves) and on-average, mesoscopic considerations (fraction of deformation from GBS, $\Gamma_{yy}^*$). Model predictions of the multi-scale deformation response, calibrated to this thermo-mechanical history and amount of sliding, then provide insights into local changes in stress fields with sliding, and the experimental observations made in [9]. Further validation of the prediction at the microscale (boundary-average sliding displacement, $\bar{\delta}_t$) in the context of the accommodation of GBS is then demonstrated in Section 7.

The model size is 200μm × 360μm × 2μm, and comprises 39 grains of mean size 40μm (Fig. 10.a). The orientations of the grains are assigned randomly, with the resulting set of orientations shown on the pole figure in Fig. 10.e. This gives rise to a material with weak texture. The 39-grain aggregate is not treated as a RVE of the material, but it enables the general creep response of polycrystals with anisotropic grains in the presence of GBS to be probed without requiring large computational cost. Note that this 39-grain configuration is the same as the thin-sliced quasi-3D geometry used in [7] (based on the approach described in [46]), where differences in macroscopic creep strain rate <15% compared to that predicted by a 216-grain full-3D RVE for the material with the same set of parameters. Discussion of appropriate RVE selection for the material can be found in Section 8.

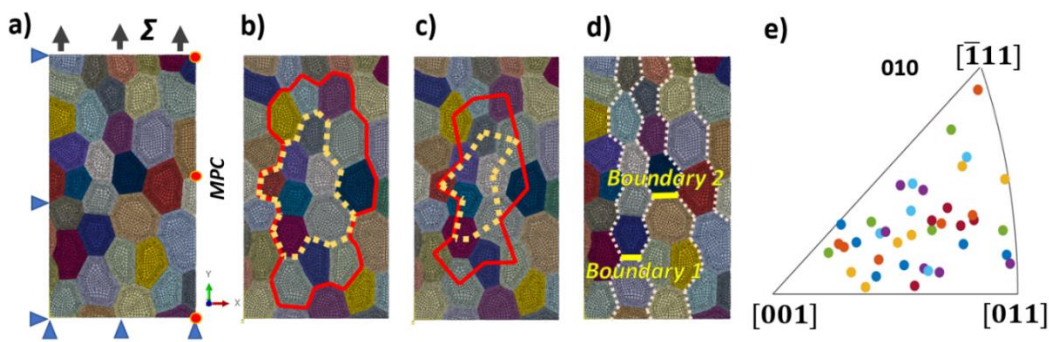

Figure 10. a) The 39-grain polycrystalline aggregate with boundary conditions. The smaller volumes of 7 grains (dashed line) and 15 grains (solid line), enclosed by b) a contour along the surrounding boundaries and c) a contour, connecting the centroids of the outer grains. d) Two boundaries, along which normal tractions are examined in Section 6.2. e) IPF (010-dir) showing the crystallographic orientations for the array (Set 1), giving weak texture.

Four meshes with different interface element lengths, $l_c$, and number of continuum linear 6-noded (C3D6) and 8-noded (C3D8) elements - i) 2.6 μm/10460 elements, ii) 1.8 μm/16417 elements, iii) 1.0 μm/35843 elements and iv) 0.75 μm/59538 elements - were investigated as part of the mesh convergence study. Mesh convergence was assumed when the difference between the predicted $E_{yy}^c$ at the end of a creep test for a finite amount of sliding was <3% with further mesh refinement. The mesh with 35843 bulk elements ($l_c$ =1.0 μm) was found to give a suitable degree of accuracy. Following from Section 5.2, the creep strain, local stress and fraction of deformation due to sliding are measured in two smaller volume aggregates within the 39-grain aggregate away from the boundaries consisting of 7 and 15 grains. This allows examination of the response of grains which are not directly affected by the details of the macroscopic boundary conditions.

A macroscopic uniaxial stress of 220 MPa is applied along the top surface of the aggregate under quasi-3D plane strain. The left-hand side of the geometry is constrained in the x-direction; the right-hand side remains straight through a multi-point constraint (MPC). The elastic interface stiffness and the value of the penalty function $P$ are identical to those used in Section 5. The calibration procedure from [6] is adopted to determine the free parameters in the micromechanical model, describing the initial state and evolution of the dislocation structure (Table 1). The calibration of the model during creep in the presence of sliding is described in the next subsection and the prediction is shown in Fig. 11.a); the short-term plasticity response is omitted for brevity. Note that solution-treated (ST) Type 316 is examined here, in which phase transformations that occur during creep cause evolution of the contributions to the internal resistance from precipitates ($\tau_p$) and solute atoms ($\tau_s$). An initial period of rapid precipitate formation and growth, followed by limited evolution leading to a constant distribution of intra- and intergranular $M_{23}C_6$ precipitates has been reported for Type 316 [9,47] over the durations considered in this section (< 300 hours). Hence, the average spacing and size of intra- and intergranular precipitates is assumed constant with $L_p = 500\ nm, r_p^m = 55\ nm;\ \lambda_p = 90\ nm, d_p = 75\ nm$ after [9,47].

Table 1. Material parameters for the CPFE model and interface elements employed in the 39-grain aggregate study describing for Type 316SS at 625°C.

| Parameter | Value | Parameter | Value |
|---|---|---|---|
| $C_{11}$ | 198 GPa* | $N_0$ | 3.0 x $10^{13}$ 1/m$^2$ |
| $C_{12}$ | 125 GPa* | $j$ | 1.75 x $10^{15}$ 1/m$^2$ |
| $C_{44}$ | 122 GPa* | $j_s$ | 8.75 x $10^{15}$ 1/m$^2$ |
| $D_c$ | 4.5 x $10^{-14}$ m$^2$/s ** | $W_c$ | 32.0 |
| $v$ | 0.39 | $\Delta$ | 8.0 |
| $G_0$ | 139 GPa | $\tau_p$ | 31 MPa |
| $b$ | 2.5 x $10^{-10}$ m | $\tau_s$ | 39 MPa |
| $\dot{\gamma}_0$ | 1.0 1/s | $\alpha_d$ | 0.35 |
| $\alpha_0$ | 1.0 | $K_t$ | $10^6$ MPa/mm |
| $K_n$ | $10^6$ MPa/mm | $\dot{\delta}_{0t}$ | $1.0 \times 10^{-26} \sim 1.0 \times 10^{-7}$ mm/s |
| $\sigma_0$ | 220 MPa | $P$ | $1 \times 10^9$ N/mm |

* Data taken from Kamaya [48]. ** Parameters from [49].

### 6.1. CALIBRATION OF POLYCRYSTALLINE MODEL IN THE PRESENCE OF SLIDING

The calibration of the polycrystalline model is based on calibration to a given macroscopic creep curve ($E_{yy}^c$ against $t$) and amount of sliding measured experimentally in the aggregate ($\Gamma_{yy}^*$). For this purpose, both the parameter controlling static recovery in the grain model ($W_c$ in Eq. 8) and the parameter controlling the amount of sliding ($\dot{\delta}_{0t}$ in Eq. 10) are modified. The procedure starts by calibrating the 5 independent model parameters, including $W_c$, which has the most significant influence on creep rate, as described in [7] in the absence of sliding. Following this, the relative sliding rate $\dot{\delta}_{0t}$ is increased until the fraction of deformation due to sliding ($\Gamma_{yy}^*$, Section 5.1.2) observed experimentally is obtained. This also modifies the overall creep response. The value of $W_c$ is then modified to provide a better fit to the creep response. The procedure is repeated until the model predictions match both the experimentally obtained creep curve and amount of sliding in the aggregate ($\Gamma_{yy}^*$). It is found that a reference sliding rate of $\dot{\delta}_{0t} = 4.5 \times 10^{-10}\ mm/s$ predicts the experimentally-recorded value of $\Gamma_{yy}^* \approx 6\%$, determined in [9] for Type 316 at 625°C and 220 MPa. This particular value of $\dot{\delta}_{0t}$ corresponds to $\phi_0 = 6.43$, obtained with respect to the reference value of macroscopic strain-rate in the absence of sliding, $\dot{E}_{yy,locked}$. The calibrated response under these conditions is shown in Fig. 11.a).

The creep response was also examined for another 9 sets of crystallographic orientations of weak texture in order to assess the effect of different detailed local configurations on the creep response and local stress state.

The same CPFE model parameters were used for all 10 sets, while the value of $\dot{\delta}_{0t}$ was modified for each set to give $\Gamma_{yy}^* \approx 6\%$. The predicted creep curves are not shown here for brevity, but differences of ±22% in creep rates compared to that given in Fig. 11.a) were found. Available experimental creep data for this material suggests that this result is within experimental scatter. In order to minimise the influence of the choice of boundary conditions on the analysed data, we also quantify the amount of deformation due to sliding $\Gamma_{yy}^*$ within the two representative volumes from Figs.10.b-c). The boundary for the group of grains in Fig. 10.b) runs along the grain boundaries. We assign half the sliding displacement experienced along these boundaries to the representative volume. The computed $\Gamma_{yy}^*$ values for the whole aggregate and the two representative volumes are compared in Fig. 11.b) across the 10 different orientation sets. This ensures that the calibrated value of $\dot{\delta}_{0t}$ to give the desired $\Gamma_{yy}^*$ is not affected by global boundary conditions. Note that the response of the aggregate at several reference sliding rates >4.5×10$^{-10}$ mm/s was also examined with the calibrated $W_c$ from Table 1 in order to assess in detail the sensitivity of local stress state. This is discussed in Section 6.2.

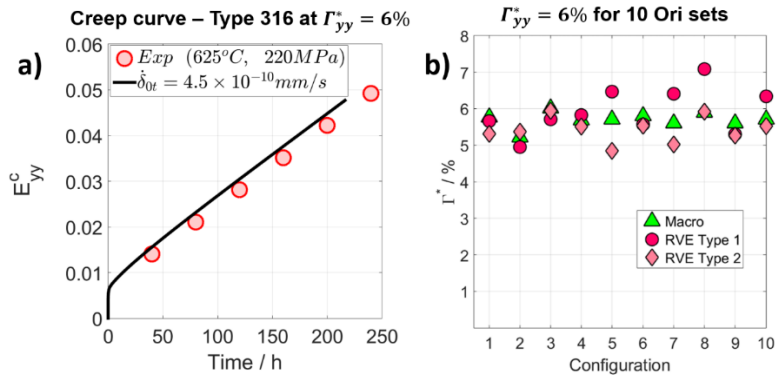

Figure 11. Calibrated creep curves at 220 MPa and 625°C for the 39-grain aggregate with the crystallographic orientations of Fig. 10.e) and experimental data from [9,50].

### 6.2 INFLUENCE OF GRAIN-BOUNDARY SLIDING ON INHOMOGENEITY OF STRESS FIELDS

The stress distributions within the smaller volumes of Figs. 10.b-c) are assessed to eliminate the effect of the global boundary conditions. Examples of heterogeneous steady-state $\sigma_{yy}$ fields within the aggregate are presented in Fig. 12. The stress distribution as a result of both sliding and mismatch due to grain misorientations resembles that described in Section 5.2. Stress concentrations near the triple lines and elevation of stress along transverse boundaries are evident with increasing $\dot{\delta}_{0t}$. The imposed local constraint prevents the opening of triple junctions. This is illustrated in Fig. 12.c) for the geometry examined in Section 6.1, but for a higher relative sliding rate of $\dot{\delta}_{ot} = 7.0 \times 10^{-9}$ mm/s. Stress concentrations are observed along the boundary, transverse to the applied load. This agrees with Section 5.3, where it was found that the local stress state depends on the constraint at the triple point. Stress concentrations along transverse boundaries near the triple junctions, resulting from sliding along adjacent boundaries, are associated with potential nucleation sites for wedge cracks, as observed by Morris and Harries [9].

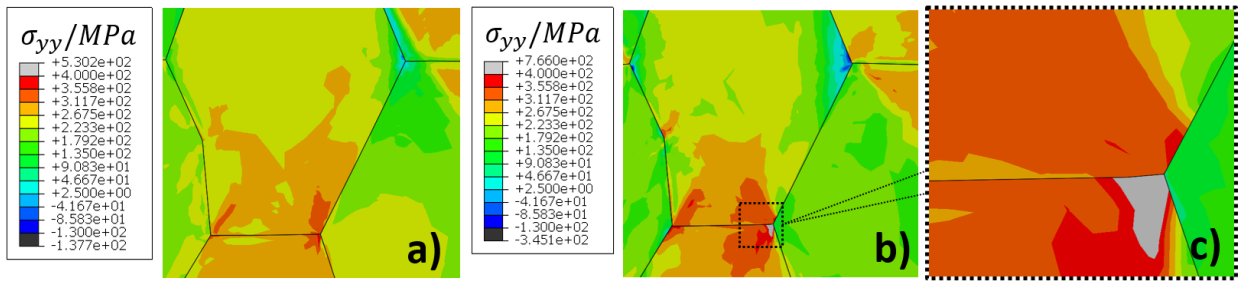

Figure 12. Contour plots of $\sigma_{yy}$ after creep for 100 hours – a) no sliding and b) $\dot{\delta}_{0t} = 7.0 \times 10^{-9}\ mm/s$ at 625°C (220 MPa) within the Type 1 representative volume. In c), a close-up view of the triple junction (b) is presented.

We now focus on the particular condition of $\Gamma_{yy}^*$ simulated in Section 6.1 ($\phi_0 = 6.43$). Examination of the predicted stress profiles along the boundaries provides more insights into the potential precursors and locations for the formation of wedge cracks. The normal stress distribution during creep along two boundaries approximately transverse to the load axis (see Fig. 10.d) within the Type 1 representative volume is examined in Fig 13. The stress distribution along these boundaries is of importance, since intergranular cavitation is primarily found on such boundaries [51]. For the scenario examined experimentally in [9] and simulated in Section 6.1 ($\phi_0 = 6.43, \dot{\delta}_{0t} = 4.5 \times 10^{-10}\ mm/s$), the local increase of $\sigma_{yy}$ in the vicinity of triple junctions is of the order of 30-40 MPa at a distance of 0.7-1.0 μm from the triple junction, as shown in Fig. 13.a)-b). For comparison, a local increase in $\sigma_{yy}$ of ~200-300 MPa at a distance of 0.7 μm from the triple line is predicted at $\phi_0 = 0.4$ ($\dot{\delta}_{0t} = 7.0 \times 10^{-9}$ mm/s, see Fig. 12). The average value of $\sigma_{yy}$ along these transverse boundaries ($\bar{\sigma}_{yy}^{GB}$) increases by ~3% for $\phi_0 = 6.43$ and ~20% for $\phi_0 = 0.4$. The effect of reference sliding rate $\dot{\delta}_{0t}$ on $\bar{\sigma}_{yy}^{GB}$ is summarised in Fig. 13.c) . These results further suggest that the standard deviation (SD($\sigma_{yy}$)) of the stress along the boundaries increases with increasing $\dot{\delta}_{0t}$. For Boundary 1, the values of (SD($\sigma_{yy}$)) are 51 ($\phi_0 = \infty$), 52.6 ($\phi_0 = 6.43$) and 61.3 ($\phi_0 = 0.4$) MPa, whereas for Boundary 2 the corresponding values are 36, 36.8 and 52.4 MPa for the same set of $\phi_0$. For the material and loading conditions examined here, the increase in heterogeneity of axial stress along the transverse boundaries is limited – with SD($\sigma_{yy}$) increasing by no more than $\approx$ 4%. Note that the stress concentrations for a given value of $\phi_0$ from Fig. 13 appear higher than those predicted in Section 5.2. This stems from the greater crystallographic mismatch and more irregular grain shapes considered here, leading to a "wedging effect" in the 39-grain aggregate.

The normal tractions along Boundary 2 are higher than those along Boundary 1. The insights provided by the results on smaller aggregates from Section 5.2 suggest that this difference is due to the different crystallographic orientations of the grains that border the boundaries (orientation set denoted O(1), see later Fig.15.a). In the remainder of this section, the influence of crystallographic orientations of neighbouring grains on the boundary normal traction is examined. The steady-state normal stress distribution along Boundaries 1 and 2, for the 10 different crystallographic orientation sets of 39 grains examined in Section 6.1, are plotted in Fig. 14. A difference of ~180 MPa arises between the average stress along a boundary for different crystallographic orientations on either side of the boundary, e.g. average normal stresses of 180 and 367 MPa along Boundary 1 are obtained for orientation sets 7 and 8, respectively.

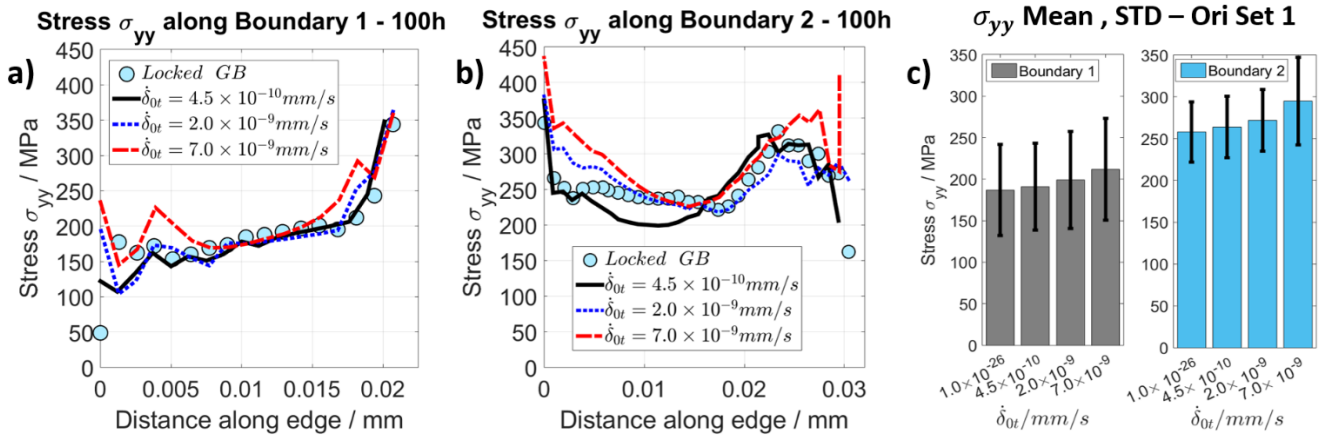

Figure 13. Line plots of $\sigma_{yy}$ with distance along x-coordinate for two transverse boundaries after creep at 625°C. The stress distributions for different amounts of sliding along a) Boundary 1 and b) Boundary 2. c) Mean ($\bar{\sigma}_{yy}^{GB}$) and standard deviation (SD($\sigma_{yy}$)) along the boundaries for four different values of reference sliding rate $\dot{\delta}_{0t}$.

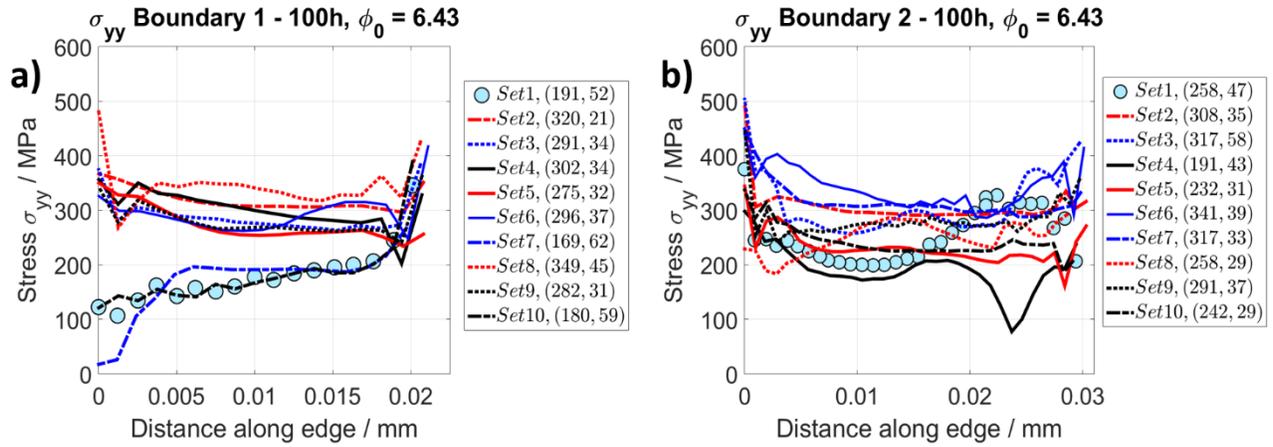

Figure 14. Line plots of $\sigma_{yy}$ along length of a) Boundary 1 and b) Boundary 2 after creep at 625°C and applied load of 220 MPa predicted for the 10 different crystallographic orientation sets examined. Values in brackets correspond to ($\bar{\sigma}_{yy}^{GB}$, SD($\sigma_{yy}$)) for the boundary computed for a given crystal orientation set.

The variation of normal stress across boundaries between grains of different crystallographic orientations agrees qualitatively with the trends from Section 5, where we demonstrate that boundaries enclosed by plastically-harder grains, with orientations away from the [001]-direction on the IPF in Fig. 10.e), carry higher normal stresses. The opposite trend is found for boundaries bordered by grains of crystallographic orientations closer to the [001]-direction, agreeing with [42,43] for other FCC materials for elastic-plastic deformation. To illustrate this further, the variation of stress along boundaries in the 39-grain aggregate is correlated with the relative crystallographic orientations of neighbouring grains. The $\sigma_{yy}$ stress fields and crystallographic orientations are examined within grains above and below Boundary 1, i.e. grains 30 and 38 in Fig. 15.a). Grains with crystallographic orientations away from the [001]-direction on the IPF in Fig.15.b) tend to carry higher stresses, e.g. grain 30 in orientation sets 8 and 9 in the $\sigma_{yy}$ plots of Fig.15.c.

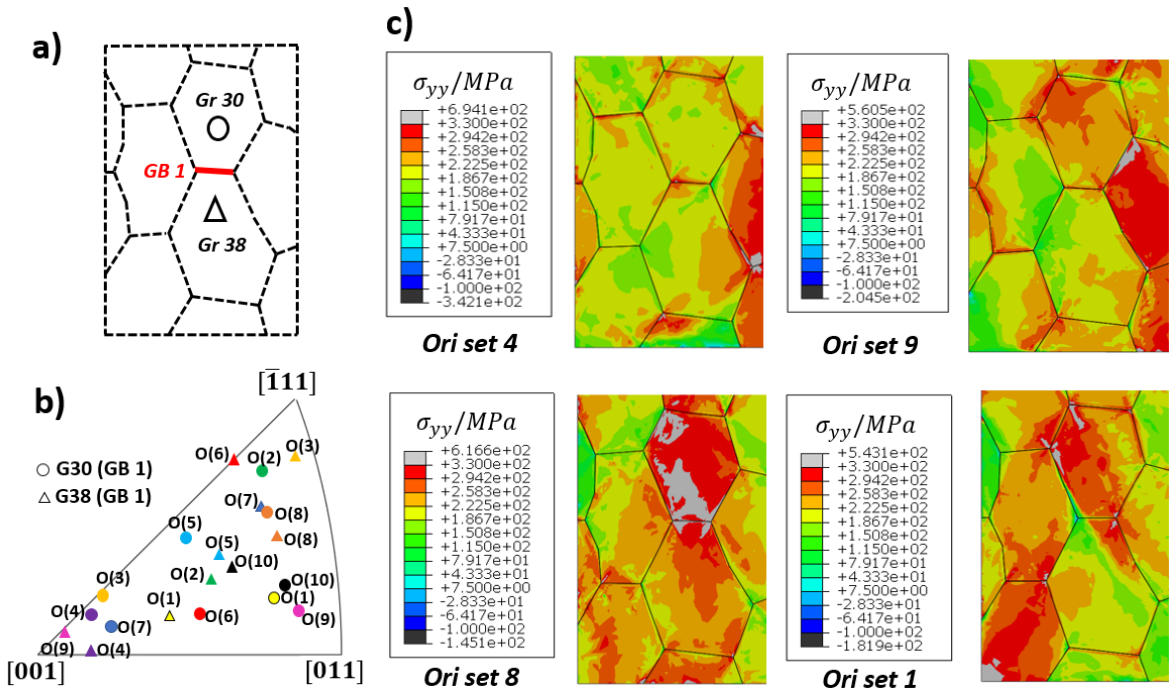

Figure 15. a) Location of Boundary 1 in the 39-grain aggregate and grains either side, namely grains 30 (o) and 38 (Δ). In b), the IPF shows the relative crystallographic orientations of grains 30 and 38 for each orientation set

(grains of the same set are presented in the same colour). In c), contour plots of the steady-state $\sigma_{yy}$ stress field are shown for three selected orientation sets indicating the higher stresses transmitted through grains of relatively harder orientations away from the [001]-direction in b).

Higher inelastic strains accumulate in grains with orientations closer to the [001]-direction, e.g. grains 30, 38 in set 4. Examination of Fig. 15 suggests that the higher normal tractions along Boundary 1 are associated with neighbouring grains of relatively harder orientations, e.g. sets 8, 9. The same trend is observed for Boundary 2. Comparison of the stress distributions along Boundaries 1 and 2 for the 10 orientation sets in the cases of no sliding and $\phi_0 = 6.43$ show that the local increase of $\sigma_{yy}$ along these boundaries within 0.7 μm of the triple lines is in the range 10-60 MPa, with an increase in mean normal stress of ~3% along these boundaries. Furthermore, the values of $\bar{\sigma}_{yy}^{GB}$ and SD($\sigma_{yy}$) averaged across the 10 sets increases from 258 and 38 MPa to 265 and 41 MPa (Boundary 1), and from 272 and 31 MPa to 275 and 37 MPa, compared to the case of no sliding. This results in an increase in $\bar{\sigma}_{yy}^{GB}$ of 2.5% and SD($\sigma_{yy}$) of 10% which is consistent with Fig. 13.c). Some of the implications of these findings are discussed in Section 8.

## 7. SLIDING ACCOMMODATION AT INTERGRANULAR PARTICLES AND EVALUATION OF MICROSCOPIC EXPERIMENTAL DATA

We now assess whether the reference sliding rate determined by calibrating the FE model is consistent with that obtained from microscopic measurements. Fig. 16 shows a comparison of the predicted average grain-boundary sliding displacement $\bar{\delta}_t$ with creep strain over all boundaries with experimental data for different austenitic stainless steels within the temperature range 600-750°C. The predictions for Type 316 are in agreement with measured values of $\bar{\delta}_t$ (0.1-0.6 μm) for Type 347 [52], Type 321 [53] and 20Cr-25Ni [54] alloys over the creep strain range 1-8%. This agreement between model predictions and experimental data for similar austenitic materials provides confidence in the accuracy of the modelling approach. These austenitic materials contain grain-boundary particles [55–57]. We now propose an approach to estimate the reference sliding rate $\dot{\delta}_{0t}$ based on the microstructure of the grain boundary, rather than calibration of the FE model, as this is found to control the sliding response [47,50,57].

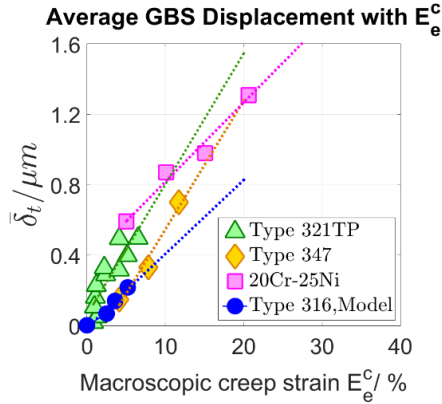

Figure 16. Comparison between model predictions for $\bar{\delta}_t$ using $\dot{\delta}_{0t} = 4.5 \times 10^{-10}\ mm/s$ and experimental data . Data from [52–54]. Linear regression fits (dashed lines) to data included.

Raj and Ashby [58,59] found that the accommodation process for grain-boundary sliding around hard particles, when grains deform by dislocation-controlled creep, is by diffusion along the particle/matrix interface. A micromechanical model is derived in [59] for $\dot{\delta}_{0t}$ with a population of particles. The model from [59] is employed to assess the effect of particle size, $d_p$, on $\dot{\delta}_{0t}$ in Type 316. $M_{23}C_6$ precipitates are the dominant intergranular particles at 625°C [57]. These precipitates can coarsen during prolonged creep exposure. We assume that the transformation of $M_{23}C_6$ precipitates is complete and therefore the volume of intergranular precipitates per unit boundary area, remains constant (i.e. $f_{VA} = d_p^3/\lambda_p^2 = const.$; $\lambda_p$ is particle spacing). Then $\dot{\delta}_{0t}$ is given as a function of the physical variables characterising the boundary, as

$$\dot{\delta}_{0t} = \frac{1.6\Omega D_L}{kTf_{VA}}\left(1 + \frac{5\delta_b D_i}{d_p D_L}\right)\sigma_0 = \alpha^p\left(1 + \frac{\beta^p}{d_p}\right)\sigma_0 \qquad (22)$$

where $D_L$ is the lattice diffusivity, $D_i$ is the particle/matrix interface diffusivity, $\delta_b$ is the boundary thickness and $\Omega$ is the atomic volume. Eq. 22 suggests that the relative sliding rate in the material decreases with increasing particle size, as larger particles require diffusional transport over greater distances to accommodate sliding. The quantities $\alpha^p$ and $\beta^p$ can be determined from independent measurements of the fundamental quantities in Eq. 22. The lattice diffusivity is known with reasonable accuracy, which allows $\alpha^p$ to be determined. Using the values of the parameters in Table 2 for Type 316 stainless steel gives $\alpha^p = 1.86 \times 10^{-16}\ mm/s.MPa$. There is, however, a great deal of uncertainty in the value of the interface diffusivity $D_i$. A suitable value can be obtained by making use of the analysis of Section 6 in which $\dot{\delta}_{0t}$ was determined by calibrating $\dot{\delta}_{0t}$ against experimental creep data. Using the calibrated value of $\dot{\delta}_{0t} = 4.5 \times 10^{-10}\ mm/s$ and $d_p$ of 75 nm as measured in [57] gives $\beta^p = 0.825\ mm$, which corresponds to $D^i = 2.05 \times 10^{-9}\ mm/s$. This diffusivity is two orders of magnitude slower than the grain boundary diffusivity given in Table 2. This magnitude of difference is consistent with the limited available data on interface diffusivity [60,61]. Once calibrated, Eq. 22 allows the effect of changing precipitate size on sliding resistance to be quantified. Thus, $\dot{\delta}_{0t}$ is dependent on the size of second-phase particles, which evolve as the material creeps. In turn, the extent of sliding can alter the evolution of local stress state along a boundary and within the grains, which can impact intergranular cavitation. This is discussed in Section 8.

Table 2. Material parameters for grain boundaries in Type 316 (625°C, $\sigma_0 = 220\ MPa$).

| Parameter | Value | Parameter | Value |
|---|---|---|---|
| $\Omega$ | 1.21 x $10^{-21}$ mm³ * | $D_b$ | 1.152 x $10^{-7}$ mm²/s § |
| $f_{VA}$ | 5.2 × $10^{-5}$ mm ** | $\delta_b$ | 5.0 x $10^{-7}$ mm * |
| $D_L$ | 6.197 x $10^{-15}$ mm²/s § | | |

* Data from [49]. ** Data from [57]. § Data interpolated from [62].

## 8. DISCUSSION

Results from Section 6 for the 39-grain aggregate suggest that, for the highest sliding rates considered ($\phi_0 = 0.4$), an increase of ~20% in the average normal traction along facets transverse to the applied uniaxial stress is expected. When replicating the experiments from [9], $\phi_0 = 6.43$ to give $\Gamma_{yy}^* \approx 6\%$, and there is ~3% predicted increase in the transverse stress and a local increase near the triple junction in the range 10-60 MPa for an applied stress of 220 MPa. The predicted increase of stress in the vicinity of triple grain junctions, where wedge cracking has been observed experimentally, suggests that the formation of wedge cracks is a result of increased local cavitation activity and conglomeration of voids near the triple junctions, rather than the decohesion mechanism proposed in [9]. This agrees with the hypothesis in [13].

    Model predictions from Section 6 further suggest that normal tractions along transverse and inclined boundaries will vary from facet to facet in polycrystalline materials, with or without grain-boundary sliding and in the absence of damage. Insights into the normal stress distribution along boundaries transverse to the applied load are of importance since coalescence of intergranular creep cavities and formation of microcracks usually occurs along these facets [63]. In Section 6, such boundaries are found to carry higher normal stresses when the neighbouring grains have plastically-harder crystallographic orientations, i.e. away from the [001]-direction on the IPF, in agreement with [42,43]. Lower stresses are transmitted across facets between plastically-softer configurations. This suggests that the rates of diffusive growth of cavities and coalescence, which is driven by the stress normal to a boundary [64], would be faster along boundaries enclosed by plastically-harder grains. A combination of experimental techniques, e.g. digital image correlation (DIC) and electron backscatter diffraction

(EBSD), could prove useful in confirming this trend. This variation of facet stress could drive the non-uniform development of intergranular creep damage observed in experiments [65].

An important finding of this study is the relative sliding observed along grain boundaries transverse to the direction of the uniaxial load. The observation that grain-boundary sliding in polycrystals, is not only a function of the grain-boundary orientation with respect to the applied load, but also the relative crystallographic orientation of the neighbouring grains, has been hypothesised in [66]. Chen [66] discussed potential implications of this phenomenon on the growth of creep cavities. Studies of cavity nucleation in the literature suggest that creep cavities could also nucleate as a result of grain-boundary sliding, see for example [67]. A recent micromechanical model for cavity nucleation in austenitic stainless steels by He and Sandström [68] expresses the nucleation rate in polycrystals as a function of the sliding rates along the boundaries. One problem with nucleation mechanisms that require grain-boundary sliding is that cavities are generally found primarily on grain boundaries transverse to the applied load [67]. From the perspective of continuum plasticity models, such boundaries do not slide and will therefore not be prone to cavitation according to this mechanism [51]. The present study demonstrates that sliding occurs along transverse boundaries and supports the hypothesis that grain-boundary sliding can be associated with cavity nucleation on these boundaries. Furthermore, the relative sliding rates in the experiments on Type 316 evaluated here are far from the freely-sliding limit, where the shear tractions along the boundaries will be fully relaxed. Therefore, the grain boundaries will carry non-zero shear tractions in addition to the normal tractions transmitted across the boundaries. The possibility of load shedding at intergranular particles also exists, e.g. Argon et al. [69]. This could impact the cavitation response, if the cavity nucleation process is stress-driven [70]. Further discussion of the factors that influence cavitation is outside the scope of this article and will be considered elsewhere.

Due to the quasi-3D nature of geometries examined, the out-of-plane component of interface sliding, present in full-3D geometries is missing and its effects on deformation and stress redistribution are not considered. GBS of full-3D arrays of grains when the grains deform by power-law creep has been examined in [18,20]. These studies suggest that the intensity of the heterogeneous stress field and increase in creep rate due to sliding, predicted in 2D aggregates, will increase in magnitude in the 3D case. This could have implications on the findings from this study for polycrystals deforming according to the CPFE model. The CPFE-interface element framework could be readily extended to study GBS in full-3D aggregates of anisotropic grains to further examine this. This poses another question on the selection of a suitable representative volume element (RVE) for the material. The aggregate is Section 6 is not treated as an RVE. It is solely adopted to illustrate some aspects of the local deformation response in the presence of sliding. An RVE for the material to study grain-boundary sliding will comprise a full-3D geometry of tessellated grains with realistic shapes (see [7] for one such geometry). Following this approach, a systematic analysis of the creep response with increasing number of grains should be carried out to determine a characteristic RVE in the presence of sliding. It is also worth highlighting the selection of an RVE in terms of the smaller cells of repeating orientation patterns from Section 5. An RVE for these geometries would be a cell, containing the least number of grains that describe the pattern. Periodic boundary conditions need to be imposed on these aggregates, in order to treat them as if they are embedded in a realistic polycrystal.

The modelling framework developed and employed in this study offers the capability of simulating the grain and grain-boundary creep deformation in polycrystalline materials, and validating the results against experimental results at various scales. Application of this framework to quantify local features of the response, such as local stress state change, which was the focus of this study, can reduce uncertainties associated with previously developed approaches which are largely based on phenomenological models. This is because of the physical basis of the lattice and GB deformation models implemented here. The modelling approach can also be extended by coupling it with phase field and atomistic models to describe in more detail the interaction between lattice slip response and the interface in the vicinity of the grain boundaries, e.g. [71]. This is beyond the scope of this paper. Nevertheless, the framework can be easily employed to study other structural alloys or pure metals, and as highlighted in preceding paragraphs, such studies should be conducted simultaneously with experimental studies at various scales which will help calibrate the model and complement the results.

# 9. CONCLUSIONS

- The developed CPFE-interface element modelling framework of creeping grains and sliding grain boundaries allows calibration and validation of the model at various scales. Approaches to inform the parameters with physical/measureable quantities, rather than empirical fitting, are proposed.
- A constraint is implemented at the triple lines to ensure that compatibility is satisfied at these locations. In the absence of this constraint, boundaries can open and errors in local stress may arise when boundaries slide.
- Relative sliding is found along interfaces transverse to the principal load direction as a result of the anisotropic deformation of the grains, leading to a mismatch in properties across a boundary. Sliding on normal interfaces, which is 1-3 orders of magnitude slower than along inclined interfaces, is required to accommodate the incompatible straining experienced either side of the boundary.
- Higher stress concentrations near triple junctions in the presence of sliding are found in configurations with a greater orientation mismatch. The normal stresses locally can be elevated by up to approximately 300 MPa at an applied stress of 220 MPa for $\Phi_0 \ll 1$. For Type 316 steel under the examined conditions, an increase in local normal stress of ~25 MPa is predicted near the triple junctions on transverse interfaces, when the fraction of the deformation due to grain-boundary sliding is 6%, with an average increase of 6.6 MPa along the boundaries.
- Grain boundaries bounded by plastically-harder grains, i.e. those with relative crystallographic orientations away from the [001]-direction, are expected to carry higher stress. The opposite is observed for plastically-softer grains closer to the [001]-direction, which accumulate higher inelastic strains.
- The characteristic rate of sliding at the interfaces is a function of the size of the grain-boundary particles. The model of Raj and Ashby [58] is used to quantify the effect of particles on the reference sliding rate $\dot{\delta}_{0t}$, which limits the uncertainty in estimating this parameter in the framework empirically.

## ACKNOWLEDGEMENTS

M.P.P. and A.C.F.C are grateful to EDF Energy for supplying the experimental data and supporting this research through PhD grant R48427/CN001. E.E. would like to acknowledge funding from the EPSRC through grant EP/L014742/1 and E.T. acknowledges funding through an EPSRC Early Career Fellowship (EP/N007239/1). The authors declare that they have no conflict of interests.

## REFERENCES

[1]. Langdon TG. Grain boundary sliding revisited: Developments in sliding over four decades. *J. Mater. Sci.* 2006;41:597–609.
[2]. Ainsworth RA. R5 procedures for assessing structural integrity of components under creep and creep–fatigue conditions. *Int. Mater. Rev.* 2006;51:107–26.
[3]. Hu J, Cocks ACF. A multi-scale self-consistent model describing the lattice deformation in austenitic stainless steels. *Int. J. Solids Struct.* Elsevier Ltd; 2016;78–79:21–37.
[4]. Hu J, Cocks ACF. Correlation Between Microstructure Evolution and Creep Properties of Polycrystalline Austenitic Stainless Steel. *Trans. SMiRT-23 Manchester, United Kingdom.* 2015.
[5]. Hu JN, Cocks ACF. Effect of creep on the Bauschinger effect in a polycrystalline austenitic stainless steel. *Scr. Mater.* 2017;128:100–4.
[6]. Petkov M, Hu J, Cocks ACF. Self-consistent modelling of cyclic loading and relaxation in austenitic 316H stainless steel. *Philos. Mag.* 2018;99:789–834.
[7]. Petkov M, Hu J, Tarleton E, Cocks ACF. Comparison of self-consistent and crystal plasticity FE approaches for modelling the high-temperature deformation of 316H austenitic stainless steel. *Int. J. Solids Struct.* 2019;171:54–80.
[8]. Garofalo F, Richmond O, Domis WF, Gemmingen F Von. Strain-time, rate-stress and rate-temperature relations during large deformations in creep. *Proc. Inst. Mech. Eng. Conf Proc.* 1963;178:1–31.
[9]. Morris DG, Harries DR. Wedge crack nucleation in Type 316 stainless steel. *J. Mater. Sci.* 1977;12:1587–97.

[10]. Gates RS. Grain boundary sliding in Type 316 austenitic steel. Part II. The creep strain and stress dependencies. *Mater. Sci. Eng.* 1977;27:115–25.
[11]. Gates RS. Grain Boundary Sliding in Type 316 Austenitic Steel. Part III: The temperature dependence. *Mater. Sci. Eng.* 1977;27:127–32.
[12]. Gates RS, Horton CAP. Grain boundary sliding in Type 316 austenitic steels. Part I: The grain size dependence. *Mater. Sci. Eng.* 1977;27:105–14.
[13]. Chen IW, Argon AS. Creep cavitation in 304 stainless steel. *Acta Metall.* 1981;29:1321–33.
[14]. Ashby MF. Boundary defects and atomistic aspects of boundary sliding and diffusional creep. *Surf. Sci.* 1972;31:498–542.
[15]. Crossman FW, Ashby MF. The non-uniform flow of polycrystals by grain-boundary sliding accommodated by power-law creep. *Acta Metall.* 1975;23:425–40.
[16]. Hsia KJ, Parks DM, Argon AS. Effects of grain boundary sliding on creep-constrained boundary cavitation and creep deformation. *Mech. Mater.* 1991;11:43–62.
[17]. Ghahremani F. Effect of grain boundary sliding on steady creep of polycrystals. *Int. J. Solids Struct.* Pergamon Press Ltd.; 1980;16:847–62.
[18]. Rodin GJ, Dib MW. Effective properties of creeping solids undergoing grain boundary sliding. Int. J. Fract. 7. 1989.
[19]. Onck P, Van Der Giessen E. Influence of microstructural variations on steady state creep and facet stresses in 2-D freely sliding polycrystals. *Int. J. Solids Struct.* 1997;34:703–26.
[20]. Anderson PM, Rice JR. Constrained creep cavitation of grain boundary facets. *Acta Metall.* 1985;33:409–22.
[21]. Needleman A, Rice JR. Plastic Creep Flow Effects in the Diffusive Cavitation of Grain Boundaries. *Acta Metall.* 1980;28:1315–32.
[22]. der Burg MWD, der Giessen E. Delaunay network modelling of creep failure in regular polycrystalline aggregates by grain boundary cavitation. *Int. J. Damage Mech.* 1994;3:111–39.
[23]. Wei YJ, Anand L. Grain-boundary sliding and separation in polycrystalline metals: Application to nanocrystalline fcc metals. *J. Mech. Phys. Solids*. 2004;52:2587–616.
[24]. Simonovski I, Cizelj L. Cohesive zone modeling of intergranular cracking in polycrystalline aggregates. *Nucl. Eng. Des.* Elsevier B.V.; 2015;283:139–47.
[25]. Pan J, Cocks ACF. Computer simulation of superplastic deformation. *Comput. Mater. Sci.* 1993;1:95–109.
[26]. Cocks ACF, Gill SPA. A variational approach to two dimensional grain growth—I. Theory. *Acta Mater.* 1996;44:4765–75.
[27]. Wilkening J, Borucki L, Sethian JA. Analysis of stress-driven grain boundary diffusion. Part I. *SIAM J. Appl. Math.* 2004;64:1839–63.
[28]. Bower AF, Wininger E. A two-dimensional finite element method for simulating the constitutive response and microstructure of polycrystals during high temperature plastic deformation. *J. Mech. Phys. Solids*. 2004;52:1289–317.
[29]. Wei Y, Bower AF, Gao H. Recoverable creep deformation and transient local stress concentration due to heterogeneous grain-boundary diffusion and sliding in polycrystalline solids. *J. Mech. Phys. Solids*. 2008;56:1460–83.
[30]. Wei Y, Bower AF, Gao H. Enhanced strain-rate sensitivity in fcc nanocrystals due to grain-boundary diffusion and sliding. *Acta Mater.* 2008;56:1741–52.
[31]. Aplin PF, Angelo DD. Dislocation-creep mechanisms in Type 316 steel. In: Wilshere B, Evans RW, editors. *Creep Fract. Eng. Mater. Struct.* Institute of Metals, London; 1990. p. 537–45.
[32]. Elmukashfi E, Cocks ACF. A theoretical and computational framework for studying creep crack growth. *Int. J. Fract.* Springer Netherlands; 2017;
[33]. Hu J, Chen B, Smith DJ, Flewitt PEJ, Cocks ACF. On the evaluation of the Bauschinger effect in an austenitic stainless steel—The role of multi-scale residual stresses. *Int. J. Plast.* Elsevier Ltd; 2016;84:203–23.
[34]. Hu J, Chen B, Smith D, Flewitt PEJ, Cocks ACF. Self-consistent model in the Local residual stress evaluation of 316H Stainless Steel. *13th Int. Conf. Fract. June 16–21*. 2013. p. 1–10.
[35]. Hu J. A theoretical study of creep deformation mechanisms of Type 316H stainless steel at high temperature. DPhil Thesis, University of Oxford; 2015.
[36]. Kocks UF, Argon AS, Ashby MF. Thermodynamics and Kinetics of Slip. *Prog. Mater. Sci.* 1975;19:1–

291.

[37]. Argon AS. Strengthening Mechanisms in Crystal Plasticity. Oxford University Press; 2008.

[38]. Kocks UF. Laws for Work-Hardening and and Low-Temperature Creep. *J. Eng. Mater. Technol.* 1976;76–86.

[39]. Ortiz M, Pandolfi A. Finite-deformation irreversible cohesive elements for three-dimensional crack-propagation analysis. *Int. J. Numer. Methods Eng.* 1999;44:1267–82.

[40]. Arora H, Tarleton E, Li-Mayer J, Charalambides MN, Lewis D. Modelling the damage and deformation process in a plastic bonded explosive microstructure under tension using the finite element method. *Comput. Mater. Sci.* Elsevier B.V.; 2015;110:91–101.

[41]. Petkov M, Chevalier M, Dean D, Cocks ACF. Creep-cyclic deformation in high temperature power plant structural materials: Informing assessment procedures in industry via multi-scale physical modelling. *Trans. SMiRT-25 Charlotte, USA*. 2019.

[42]. Chen T, Tan L, Lu Z, Xu H. The effect of grain orientation on nanoindentation behavior of model austenitic alloy Fe-20Cr-25Ni. *Acta Mater.* Elsevier Ltd; 2017;138:83–91.

[43]. Lin B, Zhao LG, Tong J. A crystal plasticity study of cyclic constitutive behaviour, crack-tip deformation and crack-growth path for a polycrystalline nickel-based superalloy. *Eng. Fract. Mech.* Elsevier Ltd; 2011;78:2174–92.

[44]. Dunne FPE, Rugg D, Walker A. Lengthscale-dependent, elastically anisotropic, physically-based hcp crystal plasticity: Application to cold-dwell fatigue in Ti alloys. *Int. J. Plast.* 2007;23:1061–83.

[45]. Shawish S El, Cizelj L, Simonovski I. Modeling grain boundaries in polycrystals using cohesive elements: Qualitative and quantitative analysis. *Nucl. Eng. Des.* 2013;261:371–81.

[46]. Lu J, Sun W, Becker A. Material characterisation and finite element modelling of cyclic plasticity behaviour for 304 stainless steel using a crystal plasticity model. *Int. J. Mech. Sci.* Elsevier; 2016;105:315–29.

[47]. Morris DG, Harries DR. Creep and rupture in Type 316 stainless steel at temperatures between 525 and 900°C Part III: Precipitation behaviour. *Met. Sci.* 1978;12:542–9.

[48]. Kamaya M. A procedure for estimating Young's modulus of textured polycrystalline materials. *Int. J. Solids Struct.* Elsevier Ltd; 2009;46:2642–9.

[49]. Ashby MF, Frost HJ. Deformation-mechanism maps: the plasticity and creep of metals and ceramics. Oxford Pergamon Press Ltd; 1982.

[50]. Morris DG, Harries DR. Creep and rupture in Type 316 stainless steel at temperatures between 525 and 900°C Part I: Creep rate. *Met. Sci.* 1978;12:525–31.

[51]. Riedel H. Fracture at High Temperatures. Springer-Verlag Berlin; 1987.

[52]. Laha K, Kyono J, Sasaki T, Kishimoto S, Shinya N. Improved creep strength and creep ductility of type 347 austenitic stainless steel through the self-healing effect of boron for creep cavitation. *Metall. Mater. Trans. A*. 2005;36:399–409.

[53]. Kishimoto S, Shinya N, Tanaka H. Grain boundary sliding and surface cracking during creep of 321 stainless steel. *Materials (Basel)*. 1987;37:289–94.

[54]. Gittins A. The kinetics of cavity growth in 20Cr/25Ni stainless steel. *J Mater Sci*. 1970;5:223–32.

[55]. Chen B, Flewitt PEJ, Smith DJ. Microstructural sensitivity of 316H austenitic stainless steel: Residual stress relaxation and grain boundary fracture. *Mater. Sci. Eng. A*. Elsevier B.V.; 2010;527:7387–99.

[56]. Martinez-Ubeda AI, Payton OD, Scott TB, Griffiths I, Younes CM. Role of long term ageing on the creep life of Type 316H austenitic stainless steel bifurcation weldments. *Proc. ASME 2016 Press. Vessel. Pip. Conf.* 2016. p. 1–7.

[57]. Morris DG, Harries DR. Creep and rupture in Type 316 stainless steel at temperatures between 525 and 900°C Part II: Rupture and ductility. *Met. Sci.* 1978;12:532–41.

[58]. Raj R, Ashby MF. Grain boundary sliding and the effects of particles on its rate. *Metall. Trans.* 1972;1937–42.

[59]. Raj R, Ashby MF. On grain boundary sliding and diffusional creep. *Metall. Trans.* 1971;2:1113–27.

[60]. Ashby MF, Centamore RM. The dragging of small oxide particles by migrating grain boundaries in copper. *Acta Metall.* 1968;16:1081–92.

[61]. Sutton AP, Balluffi RW. Interfaces in Crystalline Materials. Brook R, Cheetham A, Heuer A, Hirsch P, Marks TJ, Silcox J, et al., editors. Clarendon Press Oxford; 1995.

[62]. Perkins RA, Padgett RA, Tunali NK. Tracer diffusion of 59Fe and 51Cr in Fe-17 wt pct Cr-12 wt pct Ni

Austenitic Alloy. *Metall. Trans.* 1973;4:1665–9.
[63]. Dyson BF, Loveday MS, Rodgers MJ. Grain Boundary Cavitation Under Various States of Applied Stress. *Proc. R. Soc. A Math. Phys. Sci.* 1976;349:245–59.
[64]. Cocks ACF, Ashby MF. On creep fracture by void growth. *Prog. Mater. Sci.* 1982;27:189–244.
[65]. Pommier H, Busso EP, Morgeneyer TF, Pineau A. Intergranular damage during stress relaxation in AISI 316L-type austenitic stainless steels: Effect of carbon, nitrogen and phosphorus contents. *Acta Mater.* Elsevier Ltd; 2016;103:893–908.
[66]. Chen IW. Cavity growth on a sliding grain boundary. *Metall. Trans. A*. 1983;14:2289–93.
[67]. Kassner ME, Hayes TA. Creep cavitation in metals. *Int. J. Plast.* 2003;19:1715–48.
[68]. He J, Sandstrom R. Formation of creep cavities in austenitic stainless steels. *J. Mater. Sci.* 2016;51:6674–85.
[69]. Argon AS, Chen IW, Lau CW. Intergranular cavitation in creep: Theory and Experiments. In: Pelloux RM, Stollof N, editors. *Creep-Fatigue-Environmental Interact.* 1980. p. 46.
[70]. Raj R, Ashby MF. Intergranular fracture at elevated temperature. *Acta Metall.* 1975;23:653–66.
[71]. Zhang M, Sun K, Fang L. Influence of grain boundary activites on elastic and plastic deformation of nanocrystalline Cu as studied by phase filed and atomistic simulaiton. *Int. J. Mech. Sci.* Elsevier Ltd; 2020;187:105911.
[72]. Reinoso J, Paggi M. A consistent interface element formulation for geometrical and material nonlinearities. *Comput. Mech.* 2014;54:1569–81.

# APPENDIX A. CPFE MODEL AND INTERFACE ELEMENT IMPLEMENTATION

The CPFE model has been implemented as a user material subroutine (UMAT) within ABAQUS. The deformation gradients at the beginning and end of the time increment, $F_t$ and $F_{t+\Delta t}$ are provided and the UMAT routine computes the stress and Jacobian ($\partial\delta\sigma/\partial\delta\varepsilon$) at the end of the increment and updates state variables. An implicit scheme, using an elastic predictor and plastic corrector, following Dunne et al. [44], is employed to compute the stress through a Newton-Raphson scheme. This is described in detail in [7]. A full description of the interface element FE formation is given in [32] with only a brief summary provided here. The constitutive response of the interface is defined on the mid-surface $S_0$, between the upper and lower surfaces $S^+$, $S^-$ (Fig.A.1). Following the approach from [32], the local separations $\boldsymbol{\delta}$ ([3x1] vector) are obtained as $\boldsymbol{\delta} = \mathbf{NL_k RU} = \mathbf{B_C U}$, where $\mathbf{N}$ contains the shape functions along the parent domain, $\mathbf{L_k}$ is the displacement-separation matrix and $\mathbf{U}$ is the nodal displacement vector in global coordinates. The transformation matrix, $\mathbf{R}$, is constructed from $\mathbf{Q} = \begin{bmatrix} \bar{t}_1 & \bar{t}_2 & \bar{n} \end{bmatrix}^T$.

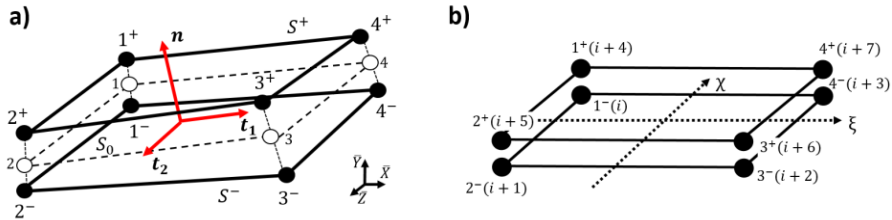

Figure A.1. Definition of interface element in 3D – a) physical and b) parent domain. An 8-noded brick element is used in 3D, defined by a normal ($\bar{n}$) and two sliding directions ($\bar{t}_1, \bar{t}_2$).

An incremental approach to compute the interface tractions is adopted. At increment $i$ and for $\dot{\delta}_n^{cr} = 0$ (no damage), the traction increments can be expressed from Eqs. 9 and 10 as

$$\Delta T_n = K_n \Delta \delta_n \; ; \; \Delta T_{tk} = \left(\Delta \delta_{tk} - B_t T_{tk}^i \Delta t\right)/\left(A_t + B_t \Delta t\right); k = 1, 2 \qquad (A.1)$$

where $\Delta t$ is the time increment, $T_{tk}^i$ are the tangential tractions at the start of the $i$-th increment, $B_t = \dot{\delta}_{0t}/\sigma_0$ and $A_t = 1/K_t$. The change in tractions over the increment is $\Delta \mathbf{T}_c = \mathbf{C}_K \Delta \boldsymbol{\delta} = \mathbf{C}_K \mathbf{B}_c \Delta \mathbf{U}$, where $\Delta \mathbf{T}_c$ is a [3x1] vector. The Jacobian matrix $\mathbf{C}_K$ for sliding along the interface are given by

$$\mathbf{C}_K = \frac{\partial \Delta \mathbf{T}_c}{\partial \Delta \boldsymbol{\delta}} = \begin{bmatrix} \frac{1}{(A_t + B_t \Delta t)} & 0 & 0 \\ 0 & \frac{1}{(A_t + B_t \Delta t)} & 0 \\ 0 & 0 & K_n \end{bmatrix} \quad (A.2)$$

Note that within a large deformation framework, the nodal coordinates of the interface element will change. The resulting deformed shape of the element will introduce additional geometrical terms within Eq. 1, as described by Ortiz and Pandolfi [39]. According to [72], the matrix $\boldsymbol{B}_c$ in Eq. 1 will evolve since the transformation matrix $\boldsymbol{R}$ will now be a function of the global displacements, $\boldsymbol{R} = \boldsymbol{R}(\boldsymbol{U})$, and not only the nodal coordinates. In this case, the full consistent tangent stiffness matrix for the interface element is given by

$$\mathbf{K}_{\text{int}} = (\partial \Psi_{\text{int}}/\partial \mathbf{U}) = \mathbf{K}_{\text{geom}} + \mathbf{K}_{\text{mat}} = \int_{\Gamma_c} (\partial \mathbf{B}_c^T/\partial \mathbf{U}) \mathbf{T}_c dS + \int_{\Gamma_c} \mathbf{B}_c^T \mathbf{C}_K \mathbf{B}_c dS \quad (A.3)$$

where $\mathbf{K}_{\text{geom}}$ is the stiffness matrix containing geometrical terms. The procedure for the determination of $\partial \mathbf{B}_c^T/\partial \mathbf{U}$ can be found in [39].

## APPENDIX B. SCHEME FOR UPDATING INTERFACE ELEMENT GEOMETRY

Under finite strains, shape functions for the reference geometry are not directly applicable to the distorted configuration. In a two-dimensional analysis, the separation rate from Eq. 10 is equal to the difference between velocities at two points directly on top of each other in Fig. B.1 (e.g. $\dot{\delta}_t = \dot{u}_{top(4*)} - \dot{u}_{bottom(1)}$).

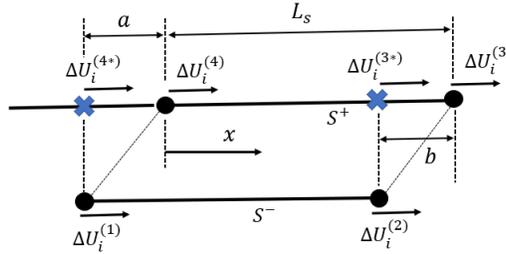

Figure B.1. Deformed interface element under 2D large deformation framework. The virtual nodes along the slave surface $S^+$ are $(3^*)$ and $(4^*)$, while actual nodes are $(3)$ and $(4)$.

In order for the shape functions of the reference configuration to be applicable to the deformed configuration, the velocity at the virtual nodes in Fig. B.1 is needed. The same linear variation velocity profile is assumed between the virtual nodes as that along the slave surface $S^+$. The incremental separation vector, evaluated by extrapolation from velocities at slave $(1, 2)$ and virtual $(3^*, 4^*)$ nodes, enclosing the reference volume for which the assumed shape function is appropriate, can be written as

$$\Delta \boldsymbol{\delta} = \left[ \Delta U_x^{(4*)} - \Delta U_x^{(1)}, \quad \Delta U_y^{(4*)} - \Delta U_y^{(1)}, \quad \Delta U_x^{(3*)} - \Delta U_x^{(2)}, \quad \Delta U_y^{(3*)} - \Delta U_y^{(2)} \right]^T \quad (B.1)$$

Based on a mixed linear interpolation-extrapolation along the local $t$-direction in 2D, the incremental displacements for nodes $(3^*)$ and $(4^*)$ are expressed in terms of the local values at nodes 3 and 4 as $\Delta U_i^{(c*)} = \Delta U_i^{(4)}\left[1 - m_c/L_s\right] + \Delta U_i^{(3)}\left[m_c/L_s\right]$, where subscript $c = 3, 4$, and $m_3 = X_1^{loc} - X_4^{loc}$ and $m_4 = L_s - (X_3^{loc} - X_2^{loc})$. This allows the separation matrix $\mathbf{L_K}$ in Appendix A to be modified as

$$\mathbf{L_K} = \begin{bmatrix} -1 & 0 & 0 & 0 & m_3/L_s & 0 & 1 - m_3/L_s & 0 \\ 0 & -1 & 0 & 0 & 0 & m_3/L_s & 0 & 1 - m_3/L_s \\ 0 & 0 & -1 & 0 & m_4/L_s & 0 & 1 - m_4/L_s & 0 \\ 0 & 0 & 0 & -1 & 0 & m_4/L_s & 0 & 1 - m_4/L_s \end{bmatrix} \quad (B.2)$$

The procedure for the 2D case is extended to 3D by extrapolating along the planar mid-surface of the 3D interface elements. The implementation is straight–forward and is omitted for brevity.

## APPENDIX C. TRIPLE LINE ELEMENTS – SHAPE FUNCTIONS

For the general 2D element, $\mathbf{A}$ is a [1x9] vector with respect to global coordinate system given by

$$\mathbf{A} = \frac{1}{2}\begin{bmatrix} n^{13}\cos(-\theta_{\Delta 1}) + n^{12}(\sin(\alpha_\Delta)/\sin(\gamma_\Delta))\cos(-\theta_{\Delta 3}) \\ -n^{13}\sin(-\theta_{\Delta 1}) - n^{12}(\sin(\alpha_\Delta)/\sin(\gamma_\Delta))\sin(-\theta_{\Delta 3}) \\ 0 \\ n^{23}(\sin(\beta_\Delta)/\sin(\gamma_\Delta))\cos(\theta_{\Delta 2}) + n^{12}(\sin(\alpha_\Delta)/\sin(\gamma_\Delta))\cos(-\theta_{\Delta 3}) \\ -n^{23}(\sin(\beta_\Delta)/\sin(\gamma_\Delta))\sin(\theta_{\Delta 2}) - n^{12}(\sin(\alpha_\Delta)/\sin(\gamma_\Delta))\sin(-\theta_{\Delta 3}) \\ 0 \\ n^{13}\cos(-\theta_{\Delta 1}) + n^{23}(\sin(\beta_\Delta)/\sin(\gamma_\Delta))\cos(\theta_{\Delta 2}) \\ -n^{13}\sin(-\theta_{\Delta 1}) - n^{23}(\sin(\beta_\Delta)/\sin(\gamma_\Delta))\sin(\theta_{\Delta 2}) \\ 0 \end{bmatrix}^T \quad (C.1)$$

where $n^{13} = -1$, $n^{23} = +1$, $n^{12} = +1$ and angles $\theta_{\Delta i}$, $\alpha_\Delta$, $\beta_\Delta$, $\gamma_\Delta$ are obtained from geometry. In the quasi-3D thin-slice geometries, $\mathbf{A}$ for the triple line element is a [1x18] vector:

$$\mathbf{A} = \frac{1}{4}\begin{bmatrix} n^{13}\cos(-\theta_{\Delta 1}) + n^{12}(\sin(\alpha_\Delta)/\sin(\gamma_\Delta))\cos(-\theta_{\Delta 3}) \\ -n^{13}\sin(-\theta_{\Delta 1}) - n^{12}(\sin(\alpha_\Delta)/\sin(\gamma_\Delta))\sin(-\theta_{\Delta 3}) \\ 0 \\ n^{23}(\sin(\beta_\Delta)/\sin(\gamma_\Delta))\cos(\theta_{\Delta 2}) + n^{12}(\sin(\alpha_\Delta)/\sin(\gamma_\Delta))\cos(-\theta_{\Delta 3}) \\ -n^{23}(\sin(\beta_\Delta)/\sin(\gamma_\Delta))\sin(\theta_{\Delta 2}) - n^{12}(\sin(\alpha_\Delta)/\sin(\gamma_\Delta))\sin(-\theta_{\Delta 3}) \\ 0 \\ n^{13}\cos(-\theta_{\Delta 1}) + n^{23}(\sin(\beta_\Delta)/\sin(\gamma_\Delta))\cos(\theta_{\Delta 2}) \\ -n^{13}\sin(-\theta_{\Delta 1}) - n^{23}(\sin(\beta_\Delta)/\sin(\gamma_\Delta))\sin(\theta_{\Delta 2}) \\ 0 \\ n^{13}\cos(-\theta'_{\Delta 1}) + n^{12}(\sin(\alpha_\Delta')/\sin(\gamma_\Delta'))\cos(-\theta'_{\Delta 3}) \\ -n^{13}\sin(-\theta'_{\Delta 1}) - n^{12}(\sin(\alpha_\Delta')/\sin(\gamma_\Delta'))\sin(-\theta'_{\Delta 3}) \\ 0 \\ n^{23}(\sin(\beta_\Delta')/\sin(\gamma_\Delta'))\cos(\theta'_{\Delta 2}) + n^{12}(\sin(\alpha_\Delta')/\sin(\gamma_\Delta'))\cos(-\theta'_{\Delta 3}) \\ -n^{23}(\sin(\beta_\Delta')/\sin(\gamma_\Delta'))\sin(\theta'_{\Delta 2}) - n^{12}(\sin(\alpha_\Delta')/\sin(\gamma_\Delta'))\sin(-\theta'_{\Delta 3}) \\ 0 \\ n^{13}\cos(-\theta'_{\Delta 1}) + n^{12}(\sin(\alpha_\Delta')/\sin(\gamma_\Delta'))\cos(-\theta'_{\Delta 3}) \\ -n^{13}\sin(-\theta'_{\Delta 1}) - n^{12}(\sin(\alpha_\Delta')/\sin(\gamma_\Delta'))\sin(-\theta'_{\Delta 3}) \\ 0 \end{bmatrix}^T \quad (C.2)$$

where the primed quantities are associated with the rear face of the triple line element.